\documentclass[12pt]{article}
\usepackage[margin=1.2in]{geometry}
\usepackage{amsmath,amssymb,amsthm}
\usepackage[authoryear]{natbib}
\usepackage{titletoc}
\usepackage[dvipsnames]{xcolor}
\usepackage{enumitem}
\usepackage{rotating}
\usepackage{float}
\usepackage{graphicx}
\usepackage{mathtools}
\usepackage{accents}
\usepackage{datetime}
\newdateformat{monthyeardate}{%
  \monthname[\THEMONTH], \THEYEAR}
\usepackage{bm}
 
\usepackage{bbm}

\usepackage[multiple]{footmisc}

\usepackage{tikz}
\usepackage{tkz-euclide}
\usepackage{pgfplots}
\pgfplotsset{compat=1.11}
\usepgfplotslibrary{fillbetween}
\usetikzlibrary{intersections}
\usetikzlibrary{fadings}
\usetikzlibrary{calc}
\pgfdeclarelayer{bg}
\pgfsetlayers{bg,main}

\newtheorem{theorem}{Theorem}
\newtheorem{proposition}{Proposition}

\newtheorem{corollary}{Corollary}

\newtheorem{fact}{Fact}
\newtheorem{definition}{Definition}

\newtheorem{example}{Example}

\theoremstyle{definition}
\newtheorem{remark}{Remark}

\usepackage{chngcntr}
\usepackage{apptools}
\AtAppendix{\counterwithin{theorem}{section}}

\newtheoremstyle{named}{}{}{\itshape}{}{\bfseries}{.}{.5em}{#1}
\theoremstyle{named}

\providecommand{\abs}[1]{\left\lvert#1 \right\rvert}
\providecommand{\norm}[1]{\left\lVert#1 \right\rVert}
\newcommand{\normo}[1]{{\left\vert\kern-0.25ex\left\vert\kern-0.25ex\left\vert #1 
    \right\vert\kern-0.25ex\right\vert\kern-0.25ex\right\vert}}
\providecommand{\pref}{\succcurlyeq}

\renewcommand{\Re}{\operatorname{\mathbb{R}}}

\DeclareMathOperator{\ca}{ca}
\DeclareMathOperator{\cl}{cl}
\renewcommand{\P}{\mathcal P}

\providecommand{\C}{\mathcal C}

\providecommand{\F}{\mathcal F}

\providecommand{\Na}{\mathbb N}

\providecommand{\Rb}{\bm{R}}

\makeatletter
\let\save@mathaccent\mathaccent
\newcommand*\if@single[3]{%
  \setbox0\hbox{${\mathaccent"0362{#1}}^H$}%
  \setbox2\hbox{${\mathaccent"0362{\kern0pt#1}}^H$}%
  \ifdim\ht0=\ht2 #3\else #2\fi
  }
\newcommand*\rel@kern[1]{\kern#1\dimexpr\macc@kerna}
\newcommand*\widebar[1]{\@ifnextchar^{{\wide@bar{#1}{0}}}{\wide@bar{#1}{1}}}
\newcommand*\wide@bar[2]{\if@single{#1}{\wide@bar@{#1}{#2}{1}}{\wide@bar@{#1}{#2}{2}}}
\newcommand*\wide@bar@[3]{%
  \begingroup
  \def\mathaccent##1##2{%
    \let\mathaccent\save@mathaccent
    \if#32 \let\macc@nucleus\first@char \fi
    \setbox\z@\hbox{$\macc@style{\macc@nucleus}_{}$}%
    \setbox\tw@\hbox{$\macc@style{\macc@nucleus}{}_{}$}%
    \dimen@\wd\tw@
    \advance\dimen@-\wd\z@
    \divide\dimen@ 3
    \@tempdima\wd\tw@
    \advance\@tempdima-\scriptspace
    \divide\@tempdima 10
    \advance\dimen@-\@tempdima
    \ifdim\dimen@>\z@ \dimen@0pt\fi
    \rel@kern{0.6}\kern-\dimen@
    \if#31
      \overline{\rel@kern{-0.6}\kern\dimen@\macc@nucleus\rel@kern{0.4}\kern\dimen@}%
      \advance\dimen@0.4\dimexpr\macc@kerna
      \let\final@kern#2%
      \ifdim\dimen@<\z@ \let\final@kern1\fi
      \if\final@kern1 \kern-\dimen@\fi
    \else
      \overline{\rel@kern{-0.6}\kern\dimen@#1}%
    \fi
  }%
  \macc@depth\@ne
  \let\math@bgroup\@empty \let\math@egroup\macc@set@skewchar
  \mathsurround\z@ \frozen@everymath{\mathgroup\macc@group\relax}%
  \macc@set@skewchar\relax
  \let\mathaccentV\macc@nested@a
  \if#31
    \macc@nested@a\relax111{#1}%
  \else
    \def\gobble@till@marker##1\endmarker{}%
    \futurelet\first@char\gobble@till@marker#1\endmarker
    \ifcat\noexpand\first@char A\else
      \def\first@char{}%
    \fi
    \macc@nested@a\relax111{\first@char}%
  \fi
  \endgroup
}
\makeatother

\providecommand{\Q}{\mathcal Q}
\DeclareMathOperator{\co}{co}
\DeclareMathOperator{\cco}{\widebar{co}}
\providecommand{\D}{\mathcal D}
\renewcommand{\L}{\mathcal L}
\providecommand{\spref}{\succ}

\providecommand{\R}{\mathcal R}
\DeclareMathOperator{\Lip}{Lip_{1}}

\providecommand{\Y}{\mathcal Y}
\providecommand{\indic}{\bm 1}

\providecommand{\Nh}{\widehat N}
\providecommand{\zt}{\tilde z}
\providecommand{\Ph}{\widehat P}
\providecommand{\CC}{\mathbf C}
\providecommand{\Ny}{N_{0}}

\usepackage{fullpage}

\allowdisplaybreaks

\usepackage[normalem]{ulem}
\usepackage[hidelinks]{hyperref}
\begin{document}

\title{Bounded arbitrage and nearly rational behavior} 

\author{Leandro Nascimento\thanks{Universidade de Bras{\' i}lia; lgnascimento@unb.br.}}

\date{Published in \textit{Economic Theory}: \url{https://doi.org/10.1007/s00199-023-01515-y}}

\maketitle


\begin{abstract} 

We establish the equivalence between a principle of almost absence of arbitrage opportunities and nearly rational decision-making. The implications of such principle are considered in the context of the aggregation of probabilistic opinions and of stochastic choice functions. In the former a bounded arbitrage principle and its equivalent form as an approximately Pareto condition are shown to bound the difference between the collective probabilistic assessment of a set of states and a linear aggregation rule on the individual assessments. In the latter we show that our general principle of limited arbitrage opportunities translates into a weakening of the McFadden–Richter axiom of stochastic rationality, and gives an upper bound for the minimum distance of a stochastic choice function to another in the class of random utility maximization models.

 \bigskip



\end{abstract}

\break 




\section{Introduction}

In a series of papers, Nau (1992, 1995a, 1995b, 2015) and Nau and McCardle (1990, 1991) advocated the use of no arbitrage conditions as a unifying principle to characterize rational behavior. They have argued that the absence of arbitrage opportunities is not only a fundamental principle underlying the modern theory of asset pricing, but can also be used to characterize a variety of forms of rational behavior. Indeed, several  expressions of rationality in economics, ranging from the axiomatic foundations of models of cardinal preferences to the correlated equilibrium outcomes in finite games, have an equivalent form as a no arbitrage condition.\footnote{Recently, Beggs (2021) has also drawn a connection between the Nau-McCardle approach and the rationalization of expenditure data \textit{\`{a} la} Afriat.}

In this paper we show that a weaker principle of bounded arbitrage opportunities characterizes models of approximate rationality for the aggregation of probabilistic opinions and in the context of stochastic choice functions. Our results build on a minimum norm duality connecting the minimum distance between two sets of probability measures to the maximum gain from arbitrage with normalized stakes. Here the possibility of arbitrage denotes a sure gain when betting on a set of states, as in the Coherence Theorem of de Finetti.\footnote{A weaker version of what is referred to as the Coherence Theorem in this paper was used in Nau and McCardle (1990, 1991) to illustrate the basic idea behind their results. Here we borrow the stronger version of the theorem as found, e.g., in Nielsen (2019, 2021). Details are given at the end of Section \ref{section: general result} below.} A consequence of our duality is that bounding the gains from arbitrage by a multiple of a positive number $\epsilon$ gives rise to weaker forms of the postulates of collective rationality in the aggregation of probabilities problem, and to a weaker version of the Axiom of Revealed Stochastic Preference, henceforth ARSP, of McFadden and Richter (1990) in the context of stochastic choice.

In the two cases we obtain the representation of a probability measure $P$ as the sum
\begin{align}\label{equation: linear aggregation with error}
	P = Q_{m}+e.
\end{align} 
Here $Q_{m}$ is a representation (a suitable linear aggregation of other measures) according to the canonical models of rational behavior. The term $e$ stands for an error  component whose length does not exceed $\epsilon$. Our characterization of nearly rational decision-making also includes a version of the two models having
\begin{align}\label{equation: intro convex combination with epsilon}
	P = (1-\epsilon)Q_{m} + \epsilon R,
\end{align}
with $Q_{m}$ as before, and for some probability measure $R$.  Like equation (\ref{equation: linear aggregation with error}), the expression in (\ref{equation: intro convex combination with epsilon}) carries a similar interpretation as a deviation from the canonical forms of rationality and can be viewed as a version of the $\epsilon$-contamination model of Huber (1964).  The standard versions of the two models correspond to the case where $\epsilon=0$. 

One such version refers to the aggregation of probabilistic assessments into a single probability measure.  This is the linear opinion pool of Stone (1961), where the elements of a set $\Q$ are interpreted as the probabilistic assessments of the odds of a set of states by the members of a group, and the probability measure $Q_{m}$ in equations (\ref{equation: linear aggregation with error}) and (\ref{equation: intro convex combination with epsilon}) is the linear averaging of the probabilities in $\Q$ with a weight vector $m$. The traditional approach to collective rational decision-making links the probabilities in $\Q$ to the probability $P$ of a social planner by means of a Pareto unanimity condition.\footnote{See Mongin (1995) and the references therein.} We show in this paper that weaker forms of the Pareto principle characterize the representations with positive $\epsilon$. Particularly, we characterize the version of an imprecise form of linear aggregation of probabilistic opinions as in (\ref{equation: intro convex combination with epsilon}), thereby handling the single-profile case of an aggregation rule due to Genest (1984).\footnote{The original formulation of Genest (1984) uses the multi-profile setting of  McConway (1981). For comparison, here we employ techniques that are different from those in Genest's paper to deal with the single-profile case.\label{footnote: Genest intro}}

We also investigate in this paper an approximate version of the random utility maximization model of Block and Marschak (1960). The exact version of this model represents a stochastic choice function as a suitably defined linear averaging of deterministic choice functions. Each element of $\Q$ stands for  a profile of choice probabilities induced by a single strict preference ordering. In this setting we give conditions ensuring that the answer to the canonical problem is perturbed proportionally to the extent of the deviation from the original condition that makes the choice probabilities in $P$ a linear averaging of the choice probabilities in the $Q\in\Q$. Our condition for an expression resembling (\ref{equation: linear aggregation with error}) is a modified version of ARSP.\footnote{The relationship (that we explore in this paper) between the standard version of ARSP and the Coherence Theorem of de Finetti also appears in the work of Clark (1996).} We also obtain a similar representation with an expression like equation (\ref{equation: intro convex combination with epsilon}) for a stochastic choice function along those same lines. This last variation of the model  was recently proposed by Apesteguia and Ballester (2021), for whom the expression in (\ref{equation: intro convex combination with epsilon}) has the component $Q_{m}$ as the stochastic choice function in a certain class predicted by the theorist, while $R$ represents unstructured randomness in the choice data that remains unexplained. Our contribution when compared to their paper is to axiomatically characterize their residual behavior representation when the class of stochastic choice functions the theorist wants to fit the data consists of the set of random utility maximization models.

This paper is organized as follows. In Section \ref{section: setting} we introduce the setting in which we work in this paper. Section \ref{section: general result} gives a general result about the proximity of the closed convex hull of two compact sets of probabilities in terms of expectations of certain functions, and relate it to the Coherence Theorem and the concept of arbitrage. In Section \ref{section: applications} we specialize our general setting in order to handle the two important applications already mentioned. While Section \ref{section: conclusion} concludes with some remarks and open problems about the questions addressed in this paper, all proofs are relegated to Appendix~A.


\section{Setting}\label{section: setting}

The set $X$ is a compact metric space with metric $d_{X}$. We denote by $\Delta(X)$ the set of all Borel probability measures on $X$. That is, each element of $\Delta(X)$ is a countably additive nonnegative measure on the Borel sets of $X$, and the measure of $X$ is normalized to one. 

For notation we write $C(X)$ to represent the vector space of all continuous functions $f\colon X\to\Re$. The vector space $C(X)$ is endowed with the supremum norm $\norm{f}_{\infty} = \sup_{x\in X}\abs{f(x)}.$ In view of continuity of the elements of $C(X)$ and compactness of $X$, the supremum can be replaced by a maximum in the definition of the norm above, as well as in the definition of the oscillation $\omega(f)$ of a function $f\in C(X)$, namely,
\[
\omega(f) = \sup_{x_{1},x_{2}\in X}[f(x_{1})-f(x_{2})].
\]
It is also convenient to define the \textit{closed} unit ball  $B_{\norm{\cdot}_{\infty}}(0,1) = \{f\in C(X):\norm{f}_{\infty}\leq 1\}$.

The set $\ca(X)$ stands throughout for the vector space of all Borel signed measures of bounded variation on $X$. Note that $\Delta(X)\subseteq \ca(X)$. The set $\ca(X)$ is endowed with the topology of weak convergence. For the purposes of this paper it suffices to notice that this is the topology where a sequence  $(R_{n})$ in $\ca(X)$ converges to $R\in\ca(X)$ if, and only if,
\[
 \lim_{n\to \infty}\int_{X} f\,dR_{n} = \int_{X} f\,dR \quad\text{ for all }f\in C(X).
\]
In the topology of weak convergence, which we refer to as the weak* (or $w*$, for short) topology, the set of probability measures $\Delta(X)$ is compact.\footnote{For this and the assertions to follow referencing the space $\ca(X)$ and its underlying topology, see Bogachev (2007, 2018).}  As a matter of notation, given $\R\subseteq \ca(X)$, the set  $\cco \R$ designates the smallest closed (in the weak* topology) and convex subset of $\ca(X)$ containing $\R$. This  will be referred to as the closed convex hull of $\R$. 

 It is also known that the weak* topology on  $\Delta(X)$ can be metrized by the so-called Kantorovich-Rubinstein metric $d_{KR}$. This metric is given by
\begin{align}
	d_{KR}(P,Q) = \sup_{f\in\Lip, \norm{f}_{\infty}\leq 1}\left(\int_{X}f\,dP -\int_{X}f\,dQ\right),\label{equation: KR metric}
\end{align}
where 
\[
	\Lip(X) = \{f\in\Re^{X}:|f(x_{1})-f(x_{2})|\leq d_{X}(x_{1},x_{2})\text{ for all }x_{1},x_{2}\in X\}.
\]

In the more conventional case where $X$ is a finite set, the topological structures of $C(X)$ and $\ca(X)$ coincide with that of a finite-dimensional Euclidean space equipped, respectively, with the maximum norm and the $\ell_{1}$ norm. Otherwise, as is well known, when the set $X$ is not finite there are in particular nonequivalent solutions, which are useful for different purposes, to the problem of giving a metric structure to the set of all probability measures on $X$.\footnote{See Gibbs and Su (2002) for a survey and comparison of the many metrics on the set $\Delta(X)$.} Here the Kantorovich-Rubinstein metric is one of many forms of quantifying the distance between two probability measures. Another important metric on the set $\Delta(X)$ is the one induced by the total variation norm $\norm{\cdot}_{1}$ on $\ca(X)$. Denoting by $\Sigma_{X}$ the collection of all Borel sets of $X$, recall that the total variation of a bounded signed measure $R\in \ca(X)$ is 
\begin{align*}
	\norm{R}_{1} = \sup \left\{\sum_{n=1}^{k}\abs{R(E_{n})}: (E_{n})_{n=1}^{k} \text{ is a partition of }X, \text{ and }E_{n}\in \Sigma_{X}\right\}.
\end{align*}
This norm induces a metric on $\Delta(X)$ as usual:
\begin{align}\label{equation: distance two measures}
	d_{1}(P,Q) = \norm{P-Q}_{1}.
\end{align}
An alternative expression for the total variation of a signed measure, which reveals a connection between the metrics $d_{KR}$ and $d_{1}$, is based on a Hahn decomposition $(X_{+},X_{-})$ of $X$. Namely, $X_{+}$ and $X_{-}$ are disjoint Borel sets of $X$ with $X = X_{+}\cup X_{-}$ and such that $R(X_{+}) = \sup_{E\in\Sigma_{X}}R(E)$ and $R(X_{-}) = \inf_{E\in\Sigma_{X}}R(E)$. Such a partition can be shown to exist and we have
\begin{align*}
	\norm{R}_{1} = R(X_{+}) - R(X_{-}).
\end{align*}
Hence we know that
\begin{align}
	d_{1}(P,Q) =  \sup_{f\in \F_{X}}\left(\int_{X}f\,dP -\int_{X}f\,dQ\right),\label{equation: total variation metric useful expression}
\end{align}
when $\F_{X}$ is the set of all measurable $f\colon X\to\Re$ with norm at most one. For the purposes of this paper we shall assume that $\F_{X} =B_{\norm{\cdot}_{\infty}}(0,1)$. Both choices of $\F_{X}$ can be used interchangeably in (\ref{equation: total variation metric useful expression}). 

We recall that in the special case where $X$ is finite and endowed with the discrete metric, the total variation and the Kantorovich-Rubinstein distance both have the same expression
\begin{align*}
	d_{1}(P,Q) = d_{KR}(P,Q) = \sum_{x\in X}\abs{P(x)-Q(x)}.
\end{align*}
This is also known as the $\ell_{1}$ distance. In general, as long as $X$ is infinite and has at least one accumulation point, the metrics $d_{1}$ and $d_{KR}$ induce different topologies on $\Delta(X)$. Without further assumptions on the metric space $X$, we have
\begin{align}
	d_{KR}(P,Q)\leq d_{1}(P,Q)\label{equation: inequality KR 1}
\end{align} 
for all $P,Q\in\Delta(X)$. The inequality in (\ref{equation: inequality KR 1}) reveals, in particular, that any pair of probability measures that are $\epsilon$-close according to the total variation distance are also $\epsilon$-close in the  Kantorovich-Rubinstein metric.\footnote{The converse implication is false as long as the metric space $(X,d_{X})$ has a countably infinite subset with an accumulation point for the same reason that in this case the topology induced by the $\ell_{1}$ norm and the topology of weak convergence are not the same. We return to this distinction in Example \ref{example: systems} in the next section.} 

In any case, unless stated otherwise, topological properties of subsets of signed measures of bounded variation on a compact metric space and the convergence of sequences of measures refer to the topology of weak convergence, and not to the notion of distance mentioned in  equation (\ref{equation: distance two measures}). Our use of the $\ell_{1}$ distance will be restricted to quantifying the length of the error term in the representations of the form expressed in~(\ref{equation: linear aggregation with error}).

\section{The general result}\label{section: general result}

To introduce our general result regarding a minimum distance duality for compact sets of probability measures, it is convenient to first provide an expression to the closure of the convex hull of such sets in terms of probability measures on them. In the next proposition we characterize the closed convex hull of a weak*-compact subset of $\Delta(X)$ as the set of all probability measures on $X$ that can be expressed as the expectation of the probabilities in that set according to some second-order probability measure. It is essentially a simple version of Choquet's theorem; see Phelps (2001).

\begin{proposition}\label{proposition: weighted average of probabilities}
		Suppose that $\R$ is a weak*-compact set of Borel probabilit\sout{ies}y measures on the compact metric space $X$. Given any  $m\in \Delta(\R)$, denote by $R_{m}$ the element of $\Delta(X)$ defined by
		\begin{align}
			R_{m}(E) = \int_{\R} R(E)\,dm \quad\text{for all Borel sets }E\subseteq X.\label{equation: expected probability}
		\end{align}
	 Then $\cco \R =  \{R_{m}\in\Delta(X): m\in \Delta(\R)\}$.
\end{proposition}

The next theorem is our minimum distance duality for the total variation distance. With the same notation as in (\ref{equation: expected probability}), the theorem characterizes the value of the problem of minimizing $\norm{P_{m_{\P}}-Q_{m_{\Q}}}_{1} $ over $(m_{\P},m_{\Q})\in\Delta(\P)\times\Delta(\Q)$, for compact sets $\P$ and $\Q$ of probabilities, with reference to the disagreement about the expected value of normalized continuous functions on $X$ between the $P\in\P$ and the $Q\in\Q$.

\begin{theorem}	\label{theorem: duality for probabilities}
		Let $\P$ and $\Q$ denote weak*-compact sets of probability measures on the compact metric space $X$. Then
	\begin{align}
		\min_{P\in\,\cco\P,Q\in\,\cco\Q}\norm{P-Q}_{1} = \sup_{\substack{f\in C(X), \norm{f}_{\infty}\leq 1}}\min_{P\in\P,Q\in \Q}\left(\int_{X}fdP -\int_{X}fdQ\right). \label{equation: minimum distance duality}
	\end{align}
\end{theorem}

A version of the theorem above appears in Luenberger (1997, p.136) exploring a different but related form of duality in order to bound the distance between two sets of continuous functions under the supremum norm. Such a form is known as the Nirenberg–Luenberger mininum distance duality. In contrast, our Theorem \ref{theorem: duality for probabilities} uses the relation between the set of signed measures on the compact metric space $X$ and the continuous linear functionals on $C(X)$ in order to prove a duality result involving two sets of probability measures endowed with the topology of weak convergence. By interpreting the minimum difference $\int_{X}f\,dP -\int_{X}f\,dQ$ as the disagreement or discrepancy of the assessment of the expected value of $f$, Theorem \ref{theorem: duality for probabilities} essentially says that the total variation distance between two closed and convex subsets of $\Delta(X)$ is the maximum discrepancy  of the assessments of expectations of certain continuous functions with respect to the probabilities in those sets. Geometrically, by viewing
\begin{align*}
	\min_{P\in\,\cco\P,Q\in\,\cco\Q}\left(\int_{X}f\,dP -\int_{X}f\,dQ\right) = \min_{P\in\,\cco \P} \int_{X}f\,dP -  \max_{Q\in\,\cco \Q} \int_{X}f\,dQ 
\end{align*}
as a scalar multiple of the distance between two parallel hyperplanes supporting, respectively, the sets $\cco \P$ and $\cco \Q$, the minimum distance between the two sets corresponds to the maximal separation. This notion is discussed at length in Dax (2006) in a finite-dimensional setting, and our Theorem \ref{theorem: duality for probabilities} becomes a special case of Theorem 13 in Dax's paper when $X$ is finite. However, the strategy of proof of that Theorem 13 relies on the compactness of the closed unit ball in finite-dimensional normed spaces, which no longer holds in the context of $C(X)$ with an infinite set $X$. To derive the expression in (\ref{equation: minimum distance duality}) we use the characterization of the total variation distance in (\ref{equation: total variation metric useful expression}) combined with a version of the minimax theorem due to Kneser (1952).

Regarding the optimization problem in (\ref{equation: minimum distance duality}), it is not difficult to check that Theorem \ref{theorem: duality for probabilities} above implies that for any two compact sets of probabilities, $\P$ and $\Q$, there are probability measures $m_{\P}\in\Delta(\P)$ and $m_{\Q}\in\Delta(\Q)$ such that 
\begin{align*}
	d_{1}(P_{m_{\P}},Q_{m_{\Q}})=  \sup_{\substack{f\in C(X), \norm{f}_{\infty}= 1}}\min_{P\in\P,Q\in \Q}\left(\int_{X}f\,dP -\int_{X}f\,dQ\right). 
\end{align*}

The identity in (\ref{equation: minimum distance duality}) is also particularly valuable because it allows us to interchange the roles of maximum and minimum in the original minimization problem. In many concrete cases arising in connection with the dualities presented in Dax and Sreedharan (1997) the expression for the minimum norm can be substantially simplified if we first calculate the minimum of the linear mapping induced by $f\in C(X)$ in (\ref{equation: minimum distance duality}). In fact, an immediate corollary of Theorem \ref{theorem: duality for probabilities} is the following version of an approximate theorem of the alternative.

\begin{corollary}\label{corollary: approximate Gordan}
	Let $\P$ and $\Q$ be as in Theorem 1, and $\epsilon>0$ be given. Then either
	\begin{description}
		\item[I.] $\int_{X}f\,dP -\int_{X}f\,dQ  >\epsilon$ for all $P\in\P$ and $Q\in \Q$ has a solution $f\in C(X)$ with $\norm{f}_{\infty}=1$,
	\end{description}
	or
	\begin{description}
		\item[II.] $d_{1}(P,Q)\leq \epsilon$ has a solution $(P,Q)\in\cco \P\times \cco \Q$,
	\end{description}
	but never both.
\end{corollary}

Corollary \ref{corollary: approximate Gordan} can be interpreted as an approximate version of Gordan's Theorem of the Alternative for probabilities. It says that either there is a strong disagreement of the expectations of normalized continuous functions on the set $X$, or else the closed convex hull of the two sets of probabilities are sufficiently close in the total variation metric of a pair of elements in those sets. For comparison, Gordan's theorem asserts that either there is some disagreement about the expectation of some $f\in C(X)$  according to the measures in $\P$ and $\Q$ in the sense that
\begin{align}
	\min_{P\in\P}\int_{X}f\,dP > \max_{Q\in\Q}\int_{X}f\,dQ,\label{equation: inequality in Gordan}
\end{align}
or else the sets $\cco \P$ and $\cco \Q$ have a nonempty intersection. This contrast becomes transparent in a finite-dimensional setting after a close inspection of the fourth column of Table 1 in Dax and Sreedharan (1997), which covers the case where $X$ is finite,  $\P$ is a singleton, and $\Q$ is the convex hull of a finite set. It should be also noted that the normalization $\norm{f}_{\infty} = 1$ cannot be omitted from system I in Corollary \ref{corollary: approximate Gordan} since the inequality in (\ref{equation: inequality in Gordan}) is equivalent to the inequality $\min_{P\in\P}\int_{X}f\,dP -\max_{Q\in\Q}\int_{X}f\,dQ  >\epsilon$ for all $f\in C(X)$ with arbitrarily large norm.

We also mention that the use of the total variation distance in Corollary \ref{corollary: approximate Gordan} can be sometimes too restrictive, so the previous corollary to our general result comes at a cost in some circumstances. We illustrate this point with the next example.

\begin{example}\label{example: systems}
 Consider a collection of approximate systems of the alternative as in Corollary \ref{corollary: approximate Gordan}, where each system is indexed by $n$. For
 \[
X = \{0\}\cup\left\{\frac{1}{k}:k\in \Na\right\}
\]
 that remains fixed and is viewed as a subset of the real line with the standard topology, in the $n$-th system we have the sets of degenerate measures $\delta_{x}$:
 \[
 \P_{n} = \{\delta_{0}\}\quad\text{and}\quad\Q_{n} = \left\{\delta_{\frac{1}{k}}:k=1,\dots,n\right\}. 
 \]
 It is not difficult to verify that in system $n$ we have
 \[\min_{P\in\cco \P_{n},Q\in\cco \Q_{n}}\norm{P-Q}_{1}=2,\]
 even though as $n$ gets larger the degenerate probability $\delta_{\frac{1}{n}}$ intuitively gets closer to $\delta_{0}$ once we take into account the standard metric structure of $X$.\footnote{This is indeed a consequence of the weak convergence of the sequence of degenerate measures $\delta_{\frac{1}{n}}$ to $\delta_{0}$.} Hence, given any $\epsilon\in(0,2)$, for every $n$ there exists a normalized continuous function $f_{n}$ such that $f_{n}(0)> f_{n}\left(\frac{1}{k}\right) + \epsilon$ for all $k=1,\dots,n$. Such functions are not difficult to construct. It is also not difficult to notice that for $n\geq\frac{1}{\epsilon}$ they cannot be Lipschitz continuous  with the Lipschitz constant equal to 1.   In view of the definition of the Kantorovich-Rubinstein metric in (\ref{equation: KR metric}), obviously either
\begin{description}
		\item[I$_{n}$.] $\int_{X}f\,dP -\int_{X}f\,dQ  >\epsilon$ for all $P\in\P_{n}$ and $Q\in \Q_{n}$ has a solution $f\in \Lip(X)$ with $\norm{f}_{\infty}\leq 1$ ,
	\end{description}
	or
	\begin{description}
		\item[II$_{n}$.] $d_{KR}(P,Q)\leq \epsilon$ has a solution $(P,Q)\in\cco \P_{n}\times \cco \Q_{n}$,
	\end{description}
	but never both. Therefore, given any $\epsilon>0$, for all $n$ sufficiently large, system I$_{n}$ does not admit a solution and system II$_{n}$ has a solution, in contrast with the behavior of the related systems under the $\ell_{1}$ metric.\qed
\end{example}

We also provide an equivalent characterization to the existence of a solution to system II in Corollary \ref{corollary: approximate Gordan}  that will be useful in the sequel.

\begin{corollary}\label{corollary: version of original epsilon alternative}
	Let $\P$ and $\Q$ be as in Theorem 1. There exists $(P,Q)\in\cco \P\times \cco \Q$ such that $d_{1}(P,Q)\leq \epsilon$ if, and only if, for all $f\in C(X)$ with $\norm{f}_{\infty}=1$,
	\begin{align}
		\min_{P\in\P}\int_{X}f\,dP\leq \max_{Q\in\Q}\int_{X}f\,dQ +\epsilon.\label{equation: inequality with epsilon separate}
	\end{align}
\end{corollary}

Corollary \ref{corollary: version of original epsilon alternative} essentially says that, in order to exist $P\in\cco \P$ and $Q\in \cco \Q$ such that $P  = Q + e$ for some error term $e\in \ca(X)$ with $\norm{e}_{1}\leq \epsilon$, the inequality in (\ref{equation: inequality with epsilon separate}) must hold for any continuous and normalized function $f$. Geometrically, it says that any continuous linear functional of unit norm on the space $\ca(X)$ cannot separate in excess of $\epsilon$ the compact and convex sets $\cco \P$ and $\cco \Q$. By applying a separating hyperplane type of argument it can be also shown that an almost identical inequality (except for a multiplicative term) can be used to characterize, given $\epsilon\in (0,1)$ and a weak* compact and convex set $\R\subseteq \Delta(X)$, the existence of probabilities $P\in\cco \P$, $Q\in\cco \Q$ and $R\in\R$ such that $P$ is the convex combination of $Q$ and $R$ with weights $1-\epsilon$ and $\epsilon$, respectively. This is done in the next proposition.

\begin{proposition}\label{proposition: characterize convex combination residual}
	Let $\P$ and $\Q$ be as in Theorem \ref{theorem: duality for probabilities}, and $\epsilon\in(0,1)$. Suppose that $\R$ is a weak* closed and convex subset of $\Delta(X)$. There exist probability measures $P\in\cco \P$, $Q\in\cco \Q$ and $R\in\R$ such that
	\begin{align}
		P = (1-\epsilon)Q+\epsilon R\label{equation: expression for general residual behavior}
	\end{align}
	 if, and only if, for all $f\in C(X)$ with $f(x)\geq 0$ for all $x\in X$ and $\norm{f}_{\infty}=1$ we have that
	\begin{align}
		\min_{P\in\P}\int_{X}f\,dP\leq (1-\epsilon)\max_{Q\in\Q}\int_{X}f\,dQ +\epsilon\max_{R\in\R}\int_{X}f\,dR .\label{equation: condition for general residual behavior}
	\end{align}
\end{proposition}

\begin{remark}
The conditions in (\ref{equation: inequality with epsilon separate}) and (\ref{equation: expression for general residual behavior}) have equivalent forms as, respectively,
	\begin{align}\label{inequality: min equivalent to max}
		\max_{P\in\P}\int_{X}f\,dP\geq \min_{Q\in\Q}\int_{X}f\,dQ -\epsilon.
	\end{align}
	and
	\begin{align*}
	\max_{P\in\P}\int_{X}f\,dP\geq (1-\epsilon)\min_{Q\in\Q}\int_{X}f\,dQ +\epsilon\min_{R\in\R}\int_{X}f\,dR.
	\end{align*} 
	These are easily verified.\qed
\end{remark}

As mentioned in the Introduction, the expressions in (\ref{equation: inequality with epsilon separate}) and (\ref{equation: condition for general residual behavior}) can both be interpreted as bounding the gains from arbitrage with normalized stakes. For concreteness we restrict attention to (\ref{equation: inequality with epsilon separate}) in its equivalent form as (\ref{inequality: min equivalent to max}), and consider a version of the Coherence Theorem of de Finetti as follows. 

Assume that $X$ has $M$ elements and the only element $P$ of $\P$ represents the prices of $M$ bets, and is thus a vector in $\Re^{M}$. Each $x\in X$ corresponds to a gamble that the individual is willing to bet on. In state $i=1,\dots, N$, the bet $x$ pays the amount $Q_{i}(x)$. We therefore write $Q_{i}$ to represent the profile of payments in state $i$ of each of those $M$ bets. Note that $\Q = \{Q_{1},\dots,Q_{N}\}$ is a subset of $\Re^{M}$. If the choice of stakes on each bet $x$ is given by  $f(x)$, then the individual receives the amount
\begin{align}\label{equation: net payoff}
	\sum_{x\in X}f(x)[Q_{i}(x)  -P(x) ]
\end{align}
in state $i$. De Finetti's Coherence Theorem can be phrased as saying that either there is a choice of stakes $f\in \Re^{M}$ with $\max_{x\in X}\abs{f(x)} = 1$ giving a sure gain (i.e., an arbitrage opportunity), meaning that the quantity in (\ref{equation: net payoff}) is positive for all $i$, or else there exists a probability measure $m$ on the set of states so that the price $P(x)$ of each gamble $x$ corresponds to its expected value $\sum_{i=1}^{N}m_{i}Q_{i}(x)$.

 According to our Theorem \ref{theorem: duality for probabilities}, if instead of restricting the sign of the expression in (\ref{equation: net payoff}) we assume that
 \begin{align}\label{equation: bounding gains finite X}
 	 \min_{i} 	\sum_{x\in X}f(x)Q_{i}(x)-\sum_{x\in X}f(x)P(x)\leq \epsilon
 \end{align}
for all $f$ with $\norm{f}_{\infty}=1$, then we know that for some $m$ in the simplex of dimension $N-1$ we have 
\begin{align}\label{equation: representation finite}
	P = \sum_{i=1}^{N}m_{i}Q_{i} +e
\end{align}
where $\norm{e}_{1}\leq \epsilon$.\footnote{In the case where $\P$ is a singleton, and $X$ and $\Q$ are finite sets, a version of this result with dual norms follows from Theorem 7.1 in Dax and Sreedharan (1997). Their result does not require that $P$ and $Q_{i}$ be elements of $\Delta(X)$, thus establishing a more general finite version of the Coherence Theorem.} The condition in (\ref{equation: bounding gains finite X}) means that (sure) gains from arbitrage obtained with normalized stakes are restricted to the interval $[0,\epsilon]$. Equation (\ref{equation: representation finite}) in its turn entails that no arbitrage principles like de Finetti's are stable. By stability we mean that when we allow for small gains from arbitrage we obtain a representation that resembles the original one except perhaps for a small error that is proportional to the small gains from arbitrage. The results to follow building on Theorem \ref{theorem: duality for probabilities} carry a similar interpretation. Since we use probabilities as primitives in this paper, we interpret conditions such as (\ref{equation: bounding gains finite X}) in terms of expectations of normalized functions (or payoffs) $f$ with respect to $P$ and the probabilities in $\Q$. Hence we read  (\ref{equation: bounding gains finite X}) like the inequality in (\ref{inequality: min equivalent to max}): for no normalized function its expected value according to $P$ is less than an $\epsilon$ of the minimum  expectation evaluated with the probabilities in $\Q$.


\section{Applications}\label{section: applications}


\subsection{Almost linear aggregation of probabilities}\label{subsection: approx linear prob}

Suppose that $\P$ is a singleton, and $\Q$ is a weak* compact set of probabilities. When the compact metric space $X$ is viewed as a set of states, the only element of $\P$, which we denote by $P$, usually has the interpretation of the probability measure expressing the assessment of a decision maker or social planner, or simply of the group, about the likelihoods of events. An element $Q$ of $\Q$ is to be interpreted as the probabilistic likelihood assessment of a member of the group of individuals about the events in $\Sigma_{X}$. Depending on the application, these members can be viewed either as members of the society, or as experts or specialists.\footnote{Compactness of the set $\Q$ can be justified when, for instance, the individuals in the group are indexed by $i\in I$, with $I$ denoting a compact metric space such as a finite set or the closed unit interval, and a continuous mapping $i\mapsto Q_{i}$ associates to each individual in the group her or his assessment of the odds of the states.}

Our first characterization of the case where the probability $P$ is sufficiently close to an element of the closed convex hull of the set $\Q$ relies on a condition that resembles the no arbitrage assumption in de Finetti’s Coherence Theorem. Two equivalent versions of such condition guarantee that $P$ is nearly represented as the linear average of the elements of $\Q$. This is our next proposition.

\begin{proposition}\label{proposition: aggregation and max min condition}
	Suppose that $\P=\{P\}\subseteq \Delta(X)$, and let $\Q$ be a nonempty and weak* compact subset of $\Delta(X)$.  Let $\epsilon>0$. There exists a countably additive (Borel) signed measure $e$ on the compact metric space $X$, and $m\in\Delta(\Q)$, such that 
	\begin{align}
		P=Q_{m}+e,\quad\text{with }\norm{e}_{1}\leq \epsilon\label{equation: expression for P}
	\end{align}
if, and only if, any of the following two equivalent conditions is satisfied.
	\begin{enumerate}
		\item[(i)] For all $f\in C(X)$ with $\norm{f}_{\infty} = 1$, $\int_{X}f\,dP\leq \max_{Q\in\Q}\int_{X} f\,dQ + \epsilon$.
		\item[(ii)] For all $f\in C(X)$ with $\norm{f}_{\infty} = 1$, $\int_{X}f\,dP\geq \min_{Q\in\Q}\int_{X} f\,dQ - \epsilon$.
	\end{enumerate}
\end{proposition} 

Proposition \ref{proposition: aggregation and max min condition} reveals that the conditions ensuring that a probability is close to a linear averaging of probabilities resemble the conditions of bounded arbitrage opportunities in the approximate version of de Finetti's theorem described at the end of Section \ref{section: general result}. For interpretation, if the continuous function $f$ in item (i) of Proposition \ref{proposition: aggregation and max min condition} represents the payoffs contingent on the states, and each probability $Q\in\Q$ encodes the opinion of an expert or better informed party about the odds of the states, the assertion in item (i) linking the elements of $\Q$ to $P$ says that, except perhaps for a small error $\epsilon$, the expected value of the normalized payoff vector $f$ according to $P$ cannot exceed the maximum expected payoff according to the opinions of the members of the group. Item (ii) can be interpreted similarly.
 
Regarding the limiting cases of Proposition \ref{proposition: aggregation and max min condition}, note first that the relevant range for the number $\epsilon$ is the interval $(0,2)$. For concreteness, for any fixed $Q\in\Q$, we can write $P = Q + e$, where $e = P-Q$ has norm $\norm{e}_{1}\leq 2$. At the same time, the assertion in item (i) in Proposition \ref{proposition: aggregation and max min condition} trivially holds for any $\epsilon\geq 2$ since
\begin{align*}
	\int_{X}f\,dP - \int_{X}f\,dQ &\leq  \norm{f}_{\infty}\norm{P-Q}_{1} \leq 2\end{align*}
for any $Q\in\Q$. And for values of $\epsilon$ close to $0$, Proposition \ref{proposition: aggregation and max min condition} has as a straightforward corollary the standard condition for linear aggregation of probabilities with reference to de Finneti's notion of coherence. This is the same as the condition in item (i) of Proposition \ref{proposition: aggregation and max min condition} with $\epsilon=0$, which is certainly equivalent to the mentioned item for all $\epsilon>0$ arbitrarily small. In fact, set $\epsilon = \frac{1}{n}$ for $n\in\Na$, and find a sequence $(m_{n})$ of measures in $\Delta(\Q)$ such that $d_{1}(P,Q_{m_{n}})\leq \frac{1}{n}$. By a standard compactness argument involving the weak$^{*}$ topology and the lower semicontinuity of the $\ell_{1}$ norm, we can find $m\in \Delta(\Q)$ such that  $P = Q_{m}$. More in general, our condition with $0<\epsilon<2$ is compatible with mild violations of coherence.\footnote{To illustrate this remark, consider Example 2 in Nielsen (2019), where  $X = \{1,2,3\}$, $P = \left(\frac{1}{3},\frac{1}{3},\frac{1}{3}\right)$, and $\Q = \{Q_{1},Q_{2}\}$, with $Q_{1} = \left(\frac{2}{3},\frac{1}{3},0\right) $ and $Q_{2} =  \left(\frac{1}{3},\frac{2}{3},0\right)$. This example reveals a violation of coherence but, as is easily verified, it is compatible with the relation in (\ref{equation: expression for P}) when $\epsilon\geq \frac{2}{3}$.}

We also characterize below the almost linear aggregation rule in (\ref{equation: expression for P}) with a version of the standard Pareto unanimity condition. This is defined next.

\begin{definition}[Condition $\CC_{\epsilon}$]\label{definition: Pareto epsilon}
	Let $\epsilon\geq 0$. We say that $P\in \Delta(X)$ and $\Q\subseteq \Delta(X)$ satisfy the condition $\CC_{\epsilon}$ when for all $f,g\in C(X)$: if $\int_{X}f\,dQ\geq \int_{X}g\,dQ$ for all $Q\in\Q$ then $\int_{X}f\,dP\geq \int_{X}g\,dP - \frac{\epsilon\cdot\omega(f-g)}{2}$. 
\end{definition}

Condition $\CC_{\epsilon}$ reduces to the standard notion of Pareto unanimity about the ranking of the payoff vectors $f$ and $g$ if $\epsilon=0$. Our notion of approximately Pareto unanimity for a positive $\epsilon$ has a correction factor of $\frac{\omega(f-g)}{2}$ in order to account for scaling effects. For instance, if we omit the oscillation of the difference $f-g$ from the last inequality in Definition \ref{definition: Pareto epsilon} its assertion would coincide with the standard form of Pareto unanimity.\footnote{To see this, just multiply each of the functions $f$ and $g$ by $n\in\Na$, and take the limit in the resulting inequalities divided by $n$ as $n\to\infty$.}

More important, while the original Pareto condition with $\epsilon=0$ would conclude with the optimality of $f$ when choosing from the set $\{f,g\}$, Definition \ref{definition: Pareto epsilon} is less demanding and requires that $f$ be nearly optimal when compared with $g$. This weaker requirement of approximate optimality is justified, for instance, as long as the decision maker with assessment $P$ of the probability of the events is uncertain about the true opinions of the members of the group, and leaves some room for probabilities that could have been omitted from the set $\Q$ and would weigh in favor of $g$. For payoff vectors $f$ and $g$ that are sufficiently close, meaning that $\norm{f-g}_{\infty}\leq 1$, a consequence of condition $\CC_{\epsilon}$ is that for the decision maker  $f$ is an $\epsilon$-optimal choice from the set $\{f,g\}$  in the sense of Radner (1980)  since in this case we also have $\int_{X}f\,dP\geq \int_{X}g\,dP -\epsilon$. Somewhat similar to the justification given in Radner (1980) for such a notion of near optimality, the parameter $\epsilon$ could reflect the difficulty the decision maker faces when gauging the preferences of the group, and is thus associated with the costs of discovering the optimal choices based on the opinions of the members of the group.

The next proposition characterizes the aggregation by nearly averaging with the weaker notion of Pareto unanimity just described.

\begin{proposition}\label{proposition: aggregation with version of Pareto}
	Suppose that $\P$, $\Q$ and $\epsilon$ are as in the statement of Proposition \ref{proposition: aggregation and max min condition}. There exist $e\in \ca(X)$ and $m\in\Delta(\Q)$ such that the relation in (\ref{equation: expression for P}) holds if, and only if, $P$ and $\Q$ satisfy condition $\CC_{\epsilon}$.
\end{proposition}

To illustrate the empirical content of Proposition \ref{proposition: aggregation with version of Pareto} we consider in the next example the aggregation of priors in an Anscombe-Aumann setting with state independent utilities.

\begin{example}
	Each individual $i=1,\dots,N$ in the group has preferences $\pref_{i}$ over the Cartesian product $\C^{M}$, where $\C$ is a nonempty convex subset of a vector space, and $M$ is a positive integer. The set $\C$ represents the set of consequences. We identify $\C^{M}$ with the set of all mappings from the $M$-element set $X$ to $\C$, and thus interpret $X$ as a set of objective states and $\C^{M}$ as the set of Anscombe-Aumann acts. Preferences of the individuals, $\pref_{i}$, and of the decision maker, $\pref_{0}$, satisfy the usual Anscombe-Aumann axioms. We assume further that preferences over consequences are the same among the individuals and the decision maker, so we let them be expressed by a non-constant and affine function $u\colon\C\to\Re$ where $u(c_{0})=0$ for some consequence $c_{0}$. We also view $c_{0}$ as the constant act with consequence $c_{0}$ in each state. Therefore, there are priors $P$ of the decision maker, and $Q_{i}$ of the members of the group such that the preference relation $\pref_{0}$ over acts $c$ is represented by $c\mapsto \sum_{x\in X}P(x)u(c(x))$, whereas $\pref_{i}$ is represented by $c\mapsto \sum_{x\in X}Q_{i}(x)u(c(x))$. In this setting, the Pareto condition ensuring that $P$ is nearly the weighted average of the probabilities $Q_{i}$ says that, if $c\pref_{i}c_{0}$ for all $i=1,\dots,N$, then $\tfrac{2}{2+\epsilon}c + \tfrac{\epsilon}{2+\epsilon}c^{\ast}\pref_{0}\tfrac{2}{2+\epsilon}c_{0} + \tfrac{\epsilon}{2+\epsilon}c_{\ast},$ where $c^{\ast}$ is the best consequence in the act $c$, and $c_{\ast}$ represents the worst consequence in $c$.		\qed
\end{example} 

In connection with the results in Section \ref{section: general result}, we now  give a single-profile version of an aggregation rule proposed by Genest (1984).\footnote{As mentioned in footnote \ref{footnote: Genest intro}, Genest's (1984) result appeared in the context of extending a result of McConway (1981), who characterized the linear pool of opinions in a multi-profile setting with two properties. The omission of one of these conditions, the so-called zero-probability property, leads to the aggregation rule of Genest in the original setting.} It consists of expressing $P$ as the convex combination of a linear averaging of the probabilities in $\Q$ and an extraneous probability measure $R\in\Delta(X)$. That is,
 \begin{align}\label{equation: aggregation a la Genest}
 	P(E) = (1-\epsilon)Q_{m}(E) + \epsilon R(E)\quad\text{for all }E\in\Sigma_{X}, 
 \end{align}
 for some $m\in\Delta(\Q)$, and $0\leq \epsilon\leq 1$. The limiting case $\epsilon=0$ corresponds to the standard linear pool of opinions, whereas the case $\epsilon=1$ posits no relation between the probability of the decision maker and the opinions of the members of the group. In general, the expression in (\ref{equation: aggregation a la Genest}) for $\epsilon\in[0,1]$ is characterized by the following version of the Pareto condition given in Definition \ref{definition: Pareto epsilon}.
 
 \begin{definition}[Condition $\CC_{\epsilon}^{\ast}$]\label{definition: new condition Pareto C}
	Let $\epsilon\geq 0$. We say that $P\in \Delta(X)$ and $\Q\subseteq \Delta(X)$ satisfy the condition $\CC_{\epsilon}^{\ast}$ when for all $f,g\in C(X)$: if $\int_{X}f\,dQ\geq \int_{X}g\,dQ$ for all $Q\in\Q$ then $\int_{X}f\,dP\geq \int_{X}g\,dP - \epsilon[\omega(f-g) - \max_{x\in X}(f(x)-g(x))]$. 
\end{definition}

A few remarks are in now order. Like condition $\CC_{\epsilon}$, the version of Pareto unanimity given in Definition \ref{definition: new condition Pareto C} says that, except perhaps for a term depending on $\epsilon$, the unanimous comparison of expected values of payoff vectors $f$ and $g$ according to the elements of $\Q$ is followed by the comparison of those payoffs according to $P$. But in contrast to the modified Pareto condition in Definition  \ref{definition: Pareto epsilon}, we have a different form for the new term when comparing expected values. Moreover, since we cannot have $\max_{x\in X}(f(x)-g(x))<0$ in Definition \ref{definition: new condition Pareto C}, for otherwise $\int_{X}f\,dQ< \int_{X}g\,dQ$, we see that condition $\CC_{\epsilon}^{\ast}$ implies condition $\CC_{2\epsilon}$. This is not surprising since any probability measure $P$ expressed as in (\ref{equation: aggregation a la Genest}) can be also written as the sum $Q_{m}+e$, where $e = \epsilon(R-Q_{m})$ has norm no greater than $2\epsilon$. 

The next proposition characterizes the aggregation rule \textit{\`{a} la} Genest with our new version of Pareto unanimity.

\begin{proposition}\label{proposition: a la Genest}
	Suppose that $\P$ and $\Q$  are as in the statement of Proposition \ref{proposition: aggregation and max min condition}. Let $\epsilon\in[0,1]$. There exist $R\in\Delta(X)$ such that the relation in (\ref{equation: aggregation a la Genest}) holds if, and only if, $P$ and $\Q$ satisfy condition $\CC_{\epsilon}^{\ast}$.
\end{proposition}

As a special case of our first aggregation result for probabilities, we also consider $P$ and the elements of $\Q$ arising as subjective probabilities in the setting of Savage (1954). Here $P$ and each $Q\in \Q$ are nonatomic.\footnote{Recall that a probability measure $R\in\Delta(X)$ is nonatomic if for all $E_{1}\in\Sigma_{X}$ with $R(E_{1})>0$ there exists $E_{2}\in \Sigma_{X}$ such that $0<R(E_{2})<R(E_{1})$.} We also assume that $\Q$ is a finite set of size $N$. According to a result due to Mongin (1995), a condition that is viewed as a form of consistency between $P$ and the set $\{Q_{1},\dots,Q_{N}\}$ when comparing the relative likelihood of events is both necessary and sufficient for $P$ to be the linear pooling of the $Q_{i}$'s. The probabilistic analogue of our modified Pareto condition can be defined as follows.

\begin{definition}[Condition $\CC^{M}_{\epsilon}$]\label{definition: Pareto binary bets}
	Let $\epsilon\geq 0$. We say that $P\in \Delta(X)$ and $\{Q_{1},\dots,Q_{N}\}\subseteq\Delta(X)$ satisfy the condition $\CC^{M}_{\epsilon}$ when for all $E_{1},E_{2}\in\Sigma_{X}$: if $Q_{i}(E_{1})\geq Q_{i}(E_{2})$ for all $i=1,\dots,N$, then $P(E_{1})\geq P(E_{2}) - \epsilon$. 
\end{definition}

Our condition $\CC^{M}_{\epsilon}$ for the case $\epsilon=0$ coincides with the consistency condition $C_{1}$ in Mongin (1995). This case is the analogue of Pareto unanimity when comparing the probability of events, or equivalently when accessing the optimality of binary bets. These are the bets on events in $\Sigma_{X}$ with a payoff of one if the event obtains, and zero otherwise. More precisely, when compared with condition $\CC_{\epsilon}$ under the assumption that $\epsilon=0$, condition $\CC^{M}_{\epsilon}$ also has an interpretation in terms of bets: replace the set $C(X)$ in Definition \ref{definition: Pareto epsilon} with the set of all indicator functions of (Borel) measurable subsets of $X$. Hence, as long as each probability in $\Q$ ranks a bet on $E_{1}$ above a bet on $E_{2}$, Definition \ref{definition: Pareto binary bets} says that according to $P$ the bet on $E_{1}$ is also ranked above the bet on $E_{2}$. The novelty in Definition \ref{definition: Pareto binary bets} with $\epsilon>0$ is to allow some room for the bet on $E_{1}$ to be worse than the bet on $E_{2}$ with respect to their expected values. But we require $P(E_{1})$ to be in the closed interval $[P(E_{2})-\epsilon,P(E_{2})]$, so the bet on $E_{1}$ cannot be too inferior when compared with the other bet.

The next proposition characterizes an almost linear aggregation of probabilities in the case of nonatomic measures. 

\begin{proposition}\label{proposition: aggregation with version of Pareto nonatomic}
	Suppose that $\P$, $\Q$ and $\epsilon$ are as in the statement of Proposition \ref{proposition: aggregation and max min condition}. Assume further that $\Q$ has size $N$, and that $P$ as well as the elements of $\Q$ are nonatomic. There exist $e\in \ca(X)$ and $m_{i}\geq 0$, for all $i=1,\dots,N$, such that $P = \sum_{i=1}^{N}m_{i}Q_{i}+e$, with $\norm{e}_{1}\leq \epsilon$, if, and only if, $P$ and $\Q$ satisfy condition $\CC^{M}_{\epsilon}$.
\end{proposition}

Our Proposition \ref{proposition: aggregation with version of Pareto nonatomic} is an extension of part of Proposition 2 in Mongin (1995). It differs from Mongin's result in two aspects. First, we assume from the outset that all the probabilities involved are nonatomic, whereas Proposition 2 in Mongin (1995) only requires that the elements of $\Q$ be nonatomic. Second, whereas Mongin (1995) obtains the exact relation $P = \sum_{i=1}^{N}m_{i}Q_{i}$ for a list of nonnegative Pareto weights $m_{i}$ whose sum is one, the expression for $P$ in our Proposition \ref{proposition: aggregation with version of Pareto nonatomic} only entails that $1-\epsilon \leq \sum_{i=1}^{N}m_{i}\leq 1+\epsilon$.\footnote{At this time we are also unaware of an example of a representation as in Proposition \ref{proposition: aggregation with version of Pareto nonatomic} never holding with $\sum_{i=1}^{N}m_{i}= 1$.} In spite of their differences, it is not difficult to see that we can recover Mongin's result as a limiting case when $\epsilon$ goes to zero.\footnote{Take $\epsilon=\frac{1}{n}$ and find a sequence of nonnegative and bounded Pareto weights $(m_{n})$ in which the representation holds accordingly for each $n$. In the limit, $P(E) = \lim_{n\to\infty}\sum_{i=1}^{N}m_{n,i}Q_{i}(E)+e_{n}(E)$, where $\norm{e_{n}}_{1}\leq \frac{1}{n}$, after passing to a convergent subsequence of $(m_{n})$ if needed.}

Finally, we can give a sharper version of Proposition \ref{proposition: aggregation with version of Pareto nonatomic} with the more general conditions for aggregation mentioned in Section \ref{section: general result}. Our next proposition characterizes a nearly linear aggregation rule for nonatomic measures with normalized Pareto weights using conditions that are similar to those in  Proposition \ref{proposition: aggregation and max min condition}.

\begin{proposition}\label{proposition: aggregation with version of Pareto nonatomic 2}
	Suppose that $\P$, $\Q$ and $\epsilon$ are as in the statement of Proposition \ref{proposition: aggregation with version of Pareto nonatomic}. There exist $e\in \ca(X)$ and $m_{i}\geq 0$, for all $i=1,\dots,N$, with $\sum_{i=1}^{N}m_{i}=1$ such that $P = \sum_{i=1}^{N}m_{i}Q_{i}+e$, where $\norm{e}_{1}\leq \epsilon$, if, and only if, any of the following two equivalent conditions is satisfied.
\begin{enumerate}
		\item[(i)] For all $E\in\Sigma_{X}$, $P(E)\leq \max_{i}Q_{i}(E)+ \frac{\epsilon}{2}$.
		\item[(ii)]  For all $E\in\Sigma_{X}$, $P(E)\geq \min_{i}Q_{i}(E)- \frac{\epsilon}{2}$.
	\end{enumerate}

\end{proposition}

\subsection{Nearly random utility maximization}\label{subsection: nearly RUM}

In the case of a finite set $X$ the results obtained in Section \ref{section: general result} can  be given the interpretation of representing a stochastic choice function as if it were nearly generated by the random utility maximization model of Block and Marschak (1960). This is done here by a suitable definition of the set $X$, and of the elements of $\P$ and $\Q$.

The setting consists of a grand set $\Y$ of alternatives of size $\Ny$, and a family $\D$ of subsets of $Y$. This family represents the choice problems the individual may choose from, and is composed of all nonempty subsets of $\Y$.\footnote{The assumption that $\D=2^{\Y}\setminus\{\emptyset \}$ is inessential and can be easily relaxed. We keep it for simplicity.} A \textit{stochastic choice function} is a real-valued mapping $P_{0}$, defined on the pairs $(y,Y)$ with $Y\in\mathcal D$ and $y\in Y$, such that
\begin{align*}
	P_{0}(y,Y)\geq 0, \quad \sum_{y\in Y}P_{0}(y,Y) = 1, \quad\text{ for all }Y\in \D.
\end{align*}
Denote by $\L$ the set of all strict preference orderings on $\Y$, namely, all those relations $\succ\,\subseteq \Y\times \Y$ that are irreflexive, transitive and satisfy the property that for any two distinct elements $y$ and $y^{\prime}$ of $\Y$ either $y\succ y^{\prime}$ or $y^{\prime}\succ y$. Let $M= \Ny\cdot 2^{\Ny-1}$, and $N= \Ny!$. Define the $M\times N$ \textit{matrix} $A$ so that each row $a(y,Y)$ has in its column corresponding to $\spref\,\in \L$ the value
\begin{align*}
		a_{\spref}(y,Y) = \begin{cases}
			1, &\text{ if }\spref\,\in S(y,Y) \\
			0, &\text{otherwise,}
		\end{cases}
	\end{align*}
where $S(y,Y) = \{\spref\,\in\L: y^{\prime}\succ y \text{ for no }y^{\prime}\in Y\}$ is the set of all strict preferences supporting the optimality of $y$ in $Y$.

The class of all stochastic choice functions that follow the random utility maximization model is the convex hull of the columns of $A$, which is known as the multiple choice polytope. To describe our approximate version of the model, we rephrase the problem in terms of two sets of probabilities as in Section \ref{section: general result}. Let $X = \{(y,Y): Y\in\D, y\in Y \}$. The set $X$ has $M$ elements and is thus a compact metric space. It is convenient to have $X$ enumerated as
\begin{align}
	X = \{(y_{1},Y_{1}),(y_{2},Y_{2}),\dots, (y_{M},Y_{M})\}.\label{equation: enumerate X trial sequence}
\end{align}
  Define $P\in\Delta(X)$ by $P(y,Y) = \frac{P_{0}(y,Y)}{2^{\Ny}-1}$, and for each $\spref\,\in\L$ the probability measure $Q_{\spref}$ on $X$ by	 $Q_{\spref}(y,Y) = \frac{a_{\spref}(y,Y)}{2^{\Ny}-1}$. The probability $P$ can be viewed as an extended lottery where $P(y,Y)$ represents the joint probability of facing the choice problem $Y$ and choosing the alternative $y\in Y$ from it. A similar interpretation applies to $Q_{\spref}$.
 	
 	The stochastic choice function $P_{0}$ belongs to the multiple choice polytope if, and only if, the sets $\P = \{P\}$ and the convex hull of the finite set $\Q = \{Q_{\spref}\in\Delta(X):\,\spref\,\in\L\}$ intersect. This is the same as saying that, for some $\pi \in\Delta(\L)$,
 	\begin{align}\label{equation: exact intersection RUM}
 		P_{0} = A\pi.
 	\end{align}
 	The probability measure $\pi$ is referred to as a stochastic preference for the obvious reasons.

 	McFadden and Richter (1990) characterize such an intersection with ARSP. We will show that, more in general, a modification of ARSP characterizes when the system of choice probabilities is $\epsilon$-close to the multiple choice polytope, and instead of the expression in (\ref{equation: exact intersection RUM}), we have
\begin{align}\label{equation: nearly RUM extended}
	P_{0} = A\pi +e
\end{align}
for some vector $e\in \Re^{M}$ with $\norm{e}_{1}\leq \epsilon$. With reference to $P$ and $\Q$, the expression in (\ref{equation: nearly RUM extended}) means that $d_{1}(P,Q_{\pi})\leq \frac{\epsilon}{2^{\Ny}-1}$ for some $Q_{\pi} = \sum_{\spref\in\L}\pi(\spref)Q_{\spref}\in\co\Q$. 

The next example, borrowed from McFadden and Richter (1990), illustrates the representation in (\ref{equation: nearly RUM extended}).

\begin{example}\label{example: MR with epsilon}
	Consider $\Y = \{y_{1},y_{2},y_{3},y_{4}\}$, and a stochastic choice function $P_{0}$ where alternatives have an equal probability of being chosen in choice problems of sizes $1$, $2$ and $4$. For choice problems of size $3$, the alternative $y_{\min}$ with the smallest index has a slightly better chance of being chosen. Denoting by $\abs{Y}$ the number of elements of $Y$,
	\begin{align*}
		P_{0}(y,Y) = \begin{cases}
			\frac{1}{\abs{Y}},&\text{ if } \abs{Y}\in\{1,2,4\}\\
			\frac{4}{10}, &\text{ if }\abs{Y}=3, y = y_{\min}\\
			\frac{3}{10}, &\text{otherwise}.
		\end{cases}
	\end{align*}
	This stochastic choice function cannot be rationalized as random utility maximization (see McFadden and Richter (1990) for the calculations). At the same time, it is not too far from a stochastic function $\tilde P_{0}$ where choices are equally likely in any choice problem. In particular,
	\begin{align*}
		P_{0}(y,Y) &=\begin{cases}
			\tilde P_{0}(y,Y) - \frac{1}{15},&\text{if }y =  y_{\min}\\
			 \tilde P_{0}(y,Y) + \frac{1}{30},&\text{otherwise}
		\end{cases}
			\end{align*}
			for a 3-element set $Y$. Hence $P_{0}$ can be represented as in (\ref{equation: nearly RUM extended}) when $\epsilon = \frac{8}{15}$. It is much less obvious, though, that $P_{0}$ can also be expressed as in (\ref{equation: nearly RUM extended}) with $\epsilon=\frac{1}{10}.$  This bound is achieved when $\tilde P_{0}(y,Y)=\sum_{\spref\in S(y,Y)}\tilde \pi(\spref)$, with $\tilde\pi $ suitably chosen.\footnote{A particular choice of  a stochastic preference  has $\tilde\pi(y_{i}\spref y_{j}\spref y_{k}\spref y_{l})  = \frac{1}{20}$ when $ijkl =1342, 2341, 3124, 3142, 3241, 4123, 4132, 4231, 4321 $, $\tilde\pi(y_{i}\spref y_{j}\spref y_{k}\spref y_{l})  = \frac{2}{20}$ when $ijkl = 1432, 2143, 2431, 3421$, and $\tilde\pi(y_{1}\spref y_{2}\spref y_{3}\spref y_{4}) = \frac{3}{20}$. This fact can be verified with the linear program mentioned in the last paragraph of Section \ref{section: conclusion} below.}	\qed
	
\end{example}

Before proceeding we need a few definitions.

\begin{definition}
	Given the enumeration of $X$ in (\ref{equation: enumerate X trial sequence}),  a tagged trial sequence $T$ is a sequence of finite and fixed length $M$ of the form
	\begin{align*}
		T =( (y_{1},Y_{1},t_{1}),(y_{2},Y_{2},t_{2}),\dots,(y_{M},Y_{M},t_{M})),
	\end{align*}
	where the $t_{i}$ are nonnegative integers. The width $w_{T} $ of a tagged trial sequence is the number $\max_{i}t_{i}-\min_{i}t_{i}$.
\end{definition}

In the original definition of McFadden and Richter (1990) and McFadden (2005), a trial sequence makes reference to a finite list of pairs $(y,Y)\in X$ where repetitions are allowed. To keep track of the greatest and the least number of repetitions in a trial sequence, we use the $t_{i}$ that are mentioned in our definition of a tagged trial sequence.\footnote{Technically, like in Section \ref{subsection: approx linear prob}, we need to make reference to the oscillation of certain functions in order to characterize the inexact form of aggregation. This is achieved by keeping track of the maximum and minimum number of repetitions using the tags.} In any case, a tagged trial sequence can be interpreted as a vector of payoffs attaching to each $(y_{i},Y_{i})\in X$ a nonnegative integer $t_{i}$. The expected payoff according to the probability $P$ on $X$ is the value $\sum_{i=1}^{M}P(y_{i},Y_{i})t_{i}$. Likewise, each  element $Q_{\spref}$ of $\Q$ also induces an expected value, which is associated with a column of $A$. In its original version in McFadden and Richter (1990), ARSP says that the expected payoff according to $P$ of any tagged trial sequence cannot exceed the maximum expected payoff computed with probabilities on $X$ induced by non-stochastic choice functions. That is, 
\begin{align}\label{equation: ARSP with probabilities}
	\sum_{i=1}^{M}P(y_{i},Y_{i})t_{i}\leq \max_{\spref\in\L}\sum_{i=1}^{M}Q_{\spref}(y_{i},Y_{i})t_{i}.
\end{align}
This is the next definition, where we use the original stochastic choice function instead of $P$, and the matrix $A$ in place of the set $\Q$.\footnote{See Border (2007) and the references therein for a discussion of ARSP.} 

\begin{definition}[ARSP]\label{definition: ARSP}
	The stochastic choice function $P_{0}$ satisfies ARSP if, for every tagged trial sequence $T$,
	\begin{align}
		\sum_{i=1}^{M}P_{0}(y_{i},Y_{i})t_{i}\leq \max_{\spref\in\L}\sum_{i=1}^{M}a_{\spref}(y_{i},Y_{i})t_{i}.\label{equation: inequality ARSP}
	\end{align}
\end{definition}

Our version of ARSP, which we will refer to as $\epsilon$-ARSP, is a weakening of the condition in (\ref{equation: inequality ARSP}). In terms of the probabilities over pairs $(y,Y)$, it consists of adjusting the right-hand side of (\ref{equation: ARSP with probabilities}) in order to accommodate deviations from the original inequality. This is similar to what is done in Corollary \ref{corollary: version of original epsilon alternative} in Section \ref{section: general result}.

\begin{definition}[$\epsilon$-ARSP]
	Let $\epsilon\geq 0$. The stochastic choice function $P_{0}$ satisfies $\epsilon$-ARSP if, for every tagged trial sequence $T$,
	\begin{align}
		\sum_{i=1}^{M}P_{0}(y_{i},Y_{i})t_{i}\leq \max_{\spref\in\L}\sum_{i=1}^{M}a_{\spref}(y_{i},Y_{i})t_{i} + \frac{w_{T}\epsilon}{2} .\label{equation: inequality epsilon ARSP}
	\end{align}
\end{definition}

In practice, $\epsilon$-ARSP is a weakening of ARSP that accommodates examples of stochastic functions that are nearly rationalizable by random utility maximization as in Example \ref{example: MR with epsilon}. We note, though, as an illustration of the content of such an axiom, that for $\epsilon<2$ it has the same implication as ARSP for non-stochastic single-valued choice functions with respect to the Weak Axiom of Revealed Preference. This observation follows from the next example.\footnote{The example can be easily extended to the more general case in which the set $\D$ of choice problems  is not comprehensive, whereby the choice function satisfies the Strong Axiom of Revealed Preference. By the strong axiom we mean that the revealed (strict) preference relation is acyclic. See McFadden and Richter (1990, pp.\ 162-165).} 

\begin{example}
	Suppose that a  deterministic choice function dictates that $y_{1}$ is chosen from $Y_{1}$, and $y_{2}\in Y_{1}$. If the weak axiom fails, for some $Y_{2}$ also containing $y_{1}$ we have $y_{2}$ as the choice from it. Now consider the tagged trial sequence $T$ where, with a slight abuse of notation, $t(y_{1},Y_{1}) = t(y_{2},Y_{2})=1$, and $t(y,Y)=0$ otherwise. Since each $\spref\,\in\L$ is transitive and irreflexive, we cannot have $a_{\spref}(y_{1},Y_{1})=a_{\spref}(y_{2},Y_{2})=1$. Hence,  using $\epsilon$-ARSP we get that, for $\epsilon<2$,
 \begin{align*}
	2 = \sum_{(y,Y)\in X}P_{0}(y,Y)t(y,Y)\leq \max_{\spref\in\L}\sum_{(y,Y)\in X}a_{\spref}(y,Y)t(y,Y) + \frac{\epsilon}{2}<2,
\end{align*}
which is impossible.	\qed 
\end{example} 

We can now state the first characterization of an approximate version of the random  utility maximization model.

\begin{proposition}\label{proposition: first epsilon random RUM}
	The stochastic choice function $P_{0}$ satisfies $\epsilon$-ARSP if, and only if, there exist $e\colon X\to\Re$ and $\pi\in\Delta(\L)$ such that 
	\begin{align}
		P_{0}(y,Y) = \sum_{\spref\in S(y,Y)}\pi(\spref) + e(y,Y),\label{equation: RUM plus e}
	\end{align}
	where $e$, viewed as an element of $\Re^{M}$, has norm $\norm{e}_{1}\leq \epsilon$.
 \end{proposition}

A second representation of nearly random utility maximization can be given with the notion of residual behavior of Apesteguia and Ballester (2021). It is a weighted combination of two terms. One stands fo the exact representation where the extended vector of choice probabilities is given as in the right-hand side of (\ref{equation: exact intersection RUM}). It is interpreted as the predicted randomness expected to be found in the choice data. The other term is the component $R_{0}\in \Re^{M}$, where $R_{0}(y,Y)\geq 0$ and $\sum_{y\in Y}R_{0}(y,Y)=1$ for all $Y\in\D$. When combined with the random utility maximization part, it leads to the expression
 \begin{align}\label{equation: nearly RUM residual behavior}
 	P_{0} = (1-\epsilon)A\pi + \epsilon R_{0}.
 \end{align}
 Comparing with (\ref{equation: nearly RUM extended}), the relation in (\ref{equation: nearly RUM residual behavior}) reveals that a fraction of the data is explained by the random utility maximization model through a stochastic preference. This is the $(1-\epsilon)A\pi$ part. The residual component, namely $\epsilon R_{0}$, is composed of what is referred to in Apesteguia and Ballester (2021) as  unstructured residual behavior. In our setting this residual refers to the portion of the data that is not explained by random utility maximization.
 
 The random utility maximization model with the residual behavior just mentioned is characterized by a version of $\epsilon$-ARSP. It is defined next.
 
 \begin{definition}[$\epsilon$-ARSP$^{\ast}$]
 	Let $\epsilon\in  [0,1]$. The stochastic choice function $P_{0}$ satisfies $\epsilon$-ARSP$^{\,\ast}$ if, for every tagged trial sequence $T$,
	\begin{align}
		\sum_{i=1}^{M}P_{0}(y_{i},Y_{i})t_{i}\leq (1-\epsilon) \max_{\spref\in\L}\sum_{i=1}^{M}a_{\spref}(y_{i},Y_{i})t_{i} + (2^{\Ny}-1)\epsilon\max_{i}t_{i}.
	\end{align}
 \end{definition}
 
 Note that any stochastic choice function represented as in (\ref{equation: nearly RUM residual behavior}) can be also expressed as in (\ref{equation: nearly RUM extended}) with $e = \epsilon(R_{0}-A\pi)$. Here, $\norm{e}_{1} = \epsilon \norm{R_{0} - A\pi}_{1}\leq 2(2^{\Ny}-1)\epsilon$. In fact, any stochastic choice function satisfying $\epsilon$-ARSP$^{\ast}$ also satisfies $\bar\epsilon$-ARSP with $\bar\epsilon = (2^{\Ny+1}-2)\epsilon$. More important, as the proposition below shows, $\epsilon$-ARSP$^{\ast}$ characterizes the representation with residual behavior for the random utility maximization model.
 
 \begin{proposition} \label{proposition: second epsilon random RUM}
 	The stochastic choice function $P_{0}$ satisfies $\epsilon$-ARSP$^{\ast}$ if, and only if, there exist $\pi\in\Delta(\L)$, and $R_{0}\in \Re^{M}$ with $R_{0}(y,Y)\geq 0$ and $\sum_{y\in Y}R_{0}(y,Y)=1$ for all $Y\in\D$, such that
 	\begin{align*}
		P_{0}(y,Y) = (1-\epsilon)\sum_{\spref\in S(y,Y)}\pi(\spref) + \epsilon R_{0}(y,Y).
	\end{align*}
 \end{proposition}


\section{Concluding remarks}\label{section: conclusion}

Many important results in economics rest on a separating hyperplane theorem. We gave approximate versions to two classical forms of rationality found in the literature. To this end, we provided characterizations of how close a pair of elements of two compact sets of probabilities are with reference to expected values of normalized functions, and also interpreted the conditions in terms of arbitrage opportunities. For the most part we turned standard arguments involving hyperplane separation into the problem of finding conditions under which two closed and convex sets are not too disjoint. We have shown that forms of nearly rational behavior obtain if, and only if, those two specific sets have points not too far from each other. The extent of departure from the classical forms of rational behavior addressed in this paper is directly proportional to the number $\epsilon$ mentioned in the text. A natural generalization of the results obtained here would be to have conditions that do not explicitly reference the number $\epsilon$.

Also in relation to the more general question studied in Section \ref{section: general result} of this paper, Hellman (2013) was the first to employ techniques similar to the ones used here to prove a related result about nearly common priors. By relying on the idea of turning the non-existence of a common prior into the possibility of strictly separating two compact and convex sets as in Samet (1998), and using the minimum norm duality theorem given in Dax (2006), Hellman (2013) shows that a condition of absence of bets with expected gains for each player greater than $\epsilon$ is equivalent to their priors having a suitably defined distance that does not exceed $\epsilon$. His setting shares a similar mathematical structure with the problem of aggregation of probabilities in the present paper, and we suspect that our results could be adapted in order to expand the framework of a finite state space in Hellman (2013). Such an extension would involve further mathematical complications and is beyond the scope of this paper.\footnote{A recent attempt at such an extension can be found in Hellman and Pint\'{e}r (2022).}

The connection between the general results presented in Section 3 above and the Coherence Theorem of de Finetti suggests that other forms of absence of arbitrage opportunities can be weakened in order to allow for approximate versions of linear pricing rules. Indeed,  Acciaio et al.\ (2022) have recently shown that approximately arbitrage-free asset prices are near a linear pricing rule. In this respect, the main difference (at least in the finite-dimensional setting) between the technical aspects of their result and the results reported in Section 3 of this paper is that while we obtain an approximate version of Gordan's Theorem of the Alternative, they establish such a version for Stiemke's Alternative. We suspect that the techniques involved in both papers can be also applied to characterize certain nonlinear pricing rules arising in the context of financial markets with frictions, such as the $\epsilon$-contamination pricing rule as described, for instance, in Araujo et al.\ (2012, 2018) and Chateauneuf and Cornet (2022).

Regarding our solution to the problem of aggregating probabilities through a nearly linear averaging procedure, we characterized the single-profile case of the aggregation rule of Genest (1984) with a weaker form of Pareto unanimity. Genest (1984) originally considered the multi-profile setting, where the aggregation rule is invariant under changes of the probabilities of the members of the group. Namely, if we refer to $I$ as the set of individuals  (a compact metric space) in the group, the version of Genest's result translates into the existence of a probability measure $m$ on $I$, and a probability measure $R$ on $X$, such that $P_{Q}(E) =(1-\epsilon) \int_{I}Q_{i}(E)\,dm(i) + \epsilon R(E)$ for all $E\in\Sigma_{X}$ and continuous profiles of probabilities $i\mapsto Q_{i}$. In the setting with a finite number of individuals it is not difficult to see that preserving the marginalization property while adding a monotonicity condition and controlling violations of the zero-probability property by $\epsilon>0$ gives a multi-profile representation of our Proposition \ref{proposition: a la Genest}.\footnote{The monotonicity condition is needed because in the original representation of Genest (1984) the weights $m(i)$ are not necessarily nonnegative except perhaps in some particular cases.} We suspect that, by adapting the arguments in Nielsen (2019), we can obtain a multi-profile version of our Proposition \ref{proposition: a la Genest} under our assumptions about $X$, the continuity of the profiles $i\mapsto Q_{i}$, and compactness of $I$. This would at the same time refine and extend Genest's (1984) characterization of the marginalization property.

Our results about the representation of stochastic choice functions relied on versions of the McFadden-Richter axiom permitting the rationalization of choice probabilities by random utility maximization. Like the McFadden-Richter axiom, the axioms $\epsilon$-ARSP and $\epsilon$-ARSP$^{\ast}$ introduced in this paper involve infinitely many restrictions. Therefore, in general we are also interested in weaker forms of the (finitely many) conditions involving the Block-Marschak (BM) polynomials used by Falmagne (1978). With the notation of Section \ref{subsection: nearly RUM}, and especially in connection with our Proposition \ref{proposition: first epsilon random RUM}, it consists of employing the double description of the stochastic choice functions $P_{0}$ following the random utility maximization model as
\begin{align}\label{equation: double description remark}
	\{A\pi: \pi\in\Delta(\L)\} = \{P_{0}: BP_{0}\geq 0, \text{$P_{0}$ is a stochastic choice function}\},
\end{align} 
for some matrix $B$ representing the BM polynomials, to derive the approximate form of the representation. By well-known results on approximate solutions to systems of linear inequalities due to Hoffman (1952) and G{\"u}ler et al.\ (1995), for a fixed $P_{0}$ there exists a constant $K>0$, depending only on $B$, such that for some $\pi\in \Delta(\L)$ we have
\begin{align*}
	\norm{P_{0}-A\pi}_{1}\leq K\norm{(BP_{0})^{-}}_{1},
\end{align*} 
where $(BP_{0})^{-}_{i} = \max\{-(BP_{0})_{i},0\}$. This shows that bounding the violations of the BM inequalities in the definition of the set in the right-hand side of (\ref{equation: double description remark}) allows us to bound the distance of a stochastic function to the family of random utility maximization models. We suspect that by using a scaled version of the parameter $\epsilon$ in the inequalities involving the BM polynomials we can also obtain an error term with $\norm{e}_{1}\leq \epsilon$  as in Proposition \ref{proposition: first epsilon random RUM}. At this point we do not have such a characterization.  Yet, like in the random utility maximization model (see McFadden and Richter, 1990, pp.\ 182-183), we can also turn the problem of deciding whether choice probabilities are nearly generated by a random utility maximization model according to (\ref{equation: RUM plus e}) into a linear programming problem. This is done by writing the problem of minimizing $\norm{P_{0} - A\pi}_{1}$ subject to $\pi\geq 0$ and $\sum_{\spref\in\L}\pi(\spref)=1$ in equivalent form as the linear program 
\begin{align*}
	&\min_{z,\pi} \sum_{i=1}^{M}z_{i}\\
	\text{s.t. }& -z_{i}\leq P_{0}(y_{i},Y_{i}) -\sum_{j=1}^{N}\pi_{j}a_{ij} \leq z_{i},\, i=1,\dots,M\\
		& \pi_{j}\geq 0,\, j=1,\dots,N\\
		&\sum_{j=1}^{N}\pi_{j}=1.
\end{align*}
Checking $\epsilon$-ARSP now involves a computational procedure that can be solved with well-known algorithms and the inspection of whether its output has $\sum_{i=1}^{M}z_{i} \leq \epsilon$.

\bigskip
\bigskip

\noindent \small{\textbf{Acknowledgements.} I am  grateful to an anonymous associate editor, and two anonymous referees for their comments and suggestions.}

\appendix
\counterwithin{fact}{subsection}

\section{Proofs}


\subsection{Proof of Proposition \ref{proposition: weighted average of probabilities}}

We first show that $R_{m}$ is well defined for each $m\in\Delta(\R)$. Since for each Borel set $E$ the indicator function $\indic_{E}$  defined on $X$, as given by
\begin{align*}
	\indic_{E}(x) =\begin{cases}
		1, &\text{if }x\in E\\
		0, &\text{if }x\notin E,
	\end{cases}
\end{align*}
 is bounded, the mapping $R\mapsto \int \indic_{E}\,dR = R(E)$ from $\R$ to $\Re$ is Borel measurable in $\R$ since it is the restriction of a Borel measurable mapping on $\Delta(X)$ (Theorem 15.13 in Aliprantis and Border, 2006). Because the mapping  $R\mapsto R(E)$ is also bounded, it follows that for any $m\in \Delta(\R)$ the value of $R_{m}(E)  $, which is nonnegative, is well defined for any Borel subset $E$ of $X$.  Obviously, $R_{m}(X) = \int_{\R}\indic_{X}\,dm = 1$. Now take any countable family $\{E_{n}:n\in\Na\}$ of pairwise disjoint elements of $\Sigma_{X}$. For each $k$, define the function $f_{k}\colon \R\to\Re$ by $f_{k}(R) = \sum_{n=1}^{k}R(E_{n})$. By the properties of the Lebesgue integral we obtain that $\int_{\R}f_{k}\,dm =\sum_{n=1}^{k}R_{m}(E_{n})$. Note that,  since $R$ is countably additive,
 \[
 \lim_{k\to\infty}f_{k}(R)= \lim_{k\to\infty}\sum_{n=1}^{k}R(E_{n}) = R(\cup_{n=1}^{\infty}E_{n})
 \]
Hence $(f_{k})$ is a bounded sequence of measurable functions that converges pointwise to the (measurable) function $f\colon\R\to\Re$ given by $f(R)  = R(\cup_{n=1}^{\infty}E_{n})$. Using the Dominated Convergence Theorem (Theorem 11.21 of Aliprantis and Border, 2006) it follows that 
\[
R_{m}(\cup_{n=1}^{\infty}E_{n}) = \int_{\R}f\,dm = \lim_{k\to\infty}\int_{\R}f_{k}\,dm =\sum_{n=1}^{\infty}R_{m}(E_{n}),
\]
thus establishing countable additivity of $R_{m}$. Therefore, $R_{m}\in \Delta(X)$.

 To complete the proof, define the set of simple probability measures on $\R$:
		\[
		\Delta_{s}(\R) = \left\{m\in\Delta(\R): m = \sum_{i=1}^{k}\lambda_{i}\delta_{R_{i}}, \exists\, \{R_{1},\dots,R_{k}\}\subseteq \R, \lambda_{i}\geq 0, \sum_{i=1}^{k}\lambda_{i}=1\right\}.
		\]
Note that $  \co \R = \{R_{m}\in\Delta(X): m\in \Delta_{s}(\R)\}\subseteq \widetilde\R $, where $\widetilde\R= \{R_{m}\in\Delta(X): m\in \Delta(\R)\}$. If $R_{m}\in \widetilde \R$, by the Density Theorem (Theorem 15.10 in Aliprantis and Border, 2006) there exists a sequence $(m_{n})$ in $\Delta_{s}(\R)$ converging to $m$. For any closed set $A\subseteq X$ the indicator function $\indic_{E}$ on $X$ is upper semicontinuous.  Hence, the function $R\mapsto R(E) = \int_{X}\indic_{E}\,dR$ is also upper semicontinuous, and so is the mapping $m\mapsto R_{m}(E)$ (see Theorem 15.5 in Aliprantis and Border, 2006).  Therefore, for any closed set $E\subseteq X$ we have
		 \begin{align}\label{equation: weak convergence of Rm}
			 \limsup_{n}R_{m_{n}}(E)\leq R_{m}(E),
		 \end{align}
 thus showing that  $R_{m_{n}}\xrightarrow{w*} R_{m}$. This establishes the inclusion $\widetilde\R\subseteq \cl_{w*}\co\R$. It now suffices to show that $\widetilde\R$ is closed. If $(R_{m_{n}})$ is a sequence in  $\widetilde\R$ converging to $R$, by weak* compactness of $\Delta(\R)$ we may assume without loss of generality that $(m_{n})$ converges to some $m\in\Delta(\R)$. By the same argument leading to (\ref{equation: weak convergence of Rm}),  we have that $(R_{m_{n}})$ converges to $R_{m}$. By uniqueness of the limit we must have $R_{m}=R$. We conclude that
  \begin{align*}
		 	\cco \R = \cl_{w*}\co\R\subseteq \widetilde \R\subseteq  \cl_{w*}\co\R,
	\end{align*} 
and thus $\cco\R = \{R_{m}\in\Delta(X): m\in \Delta(\R)\}$.

		
\subsection{Proof of Theorem \ref{theorem: duality for probabilities}}

The total variation norm of the difference of two probability measures $P$ and $Q$ can be written as the composition of the  mappings $F\colon\Delta(X)\times \Delta(X)\to \ca(X)$, as given by $F(P,Q) = P-Q$, and $\norm{\cdot}_{1}$. The latter is viewed as a real-valued function on $\ca(X)$. The Cartesian product $\Delta(X)\times \Delta(X)$ is equipped with the product topology. Hence the mapping $F$ is continuous. Since $\norm{\cdot}_{1}$, as the dual norm of the sup norm on $C(X)$, is a lower semicontinuous function when $\ca(X)$ is endowed with the topology of weak convergence (see Lemma 6.22 in Aliprantis and Border, 2006), we conclude that $\norm{P-Q}_{1} = \norm{F(P,Q)}_{1}$ defines a lower semicontinuous function on $\Delta(X)\times \Delta(X)$. In particular, the minimum of $\norm{P-Q}_{1}$, with $P$ ranging over $\cco\P$ and $Q$ ranging over $\cco \Q$, is attained. This assertion follows from compactness of the set $\cco\P\times\cco\Q$, denoted $\R$, and the lower semicontinuity of $\norm{F(P,Q)}_{1}$. By a similar argument, for each function $f\in C(X)$, the minimum of $\int_{X}fdP -\int_{X}fdQ$ with $P\in\P$ and $Q\in\Q$ is also attained. Using the definition of the $\ell_{1}$ distance in Section \ref{section: setting}, or simply the characterization of the total variation norm in terms of its duality with the sup norm, we obtain that
		\begin{align*}
			\min_{P\in\,\cco\P,Q\in\,\cco\Q}\norm{P-Q}_{1} =\min_{\Rb\,\in\,\R}\sup_{f\in B_{\norm{\cdot}_{\infty}}(0,1)}G(f,\Rb),
		\end{align*} 
	where $G\colon B_{\norm{\cdot}_{\infty}}(0,1)\times \R\to\Re$ is the continuous function given by $G(f,\Rb) = \int_{X}fdP - \int_{X}fdQ$ for $\Rb = (P,Q)$. Note that $G$ is affine in each variable, the closed unit ball $ B_{\norm{\cdot}_{\infty}}(0,1)$ of $C(X)$ is convex, and $\R$ is a compact and convex subset of $\Delta(X)\times\Delta(X)$. By the minimax theorem (Theorem N$^{\prime}$ in Kneser (1952), or Corollary 3.3 in Sion (1958)),	
	\begin{align}
			\min_{\Rb\,\in\,\R}\sup_{f\in B_{\infty}(0,1)}G(f,\Rb) &=\sup_{f\in B_{\norm{\cdot}_{\infty}}(0,1)} \min_{\Rb\,\in\,\R}G(f,\Rb) \notag\\
			& =  \sup_{f\in B_{\norm{\cdot}_{\infty}}(0,1)}\min_{P\in\cco \P,Q\in\cco \Q}\left(\int_{X}f\,dP - \int_{X}f\,dQ\right)\notag\\
			& = \sup_{f\in B_{\norm{\cdot}_{\infty}}(0,1)}\min_{P\in \P,Q\in \Q}\left(\int_{X}f\,dP - \int_{X}f\,dQ\right). \label{equation: equality in sup min}
		\end{align} 
	Note that the  equality in (\ref{equation: equality in sup min}) is easily verified since given any two sequences $(P_{n})$ and $(Q_{n})$ in $\co\P$ and $\co \Q$, respectively, we have that
	\begin{align*}
			\min_{P\in \P,Q\in \Q}\int_{X}f\,dP - \int_{X}f\,dQ\leq \int_{X}f\,dP_{n} - \int_{X}f\,dQ_{n}
		\end{align*}
		because of linearity of the linear funcional induced by $f$ in the space $\ca(X)$, so by weak convergence,  upon letting $P_{n}\xrightarrow{w*} P$ and $Q_{n}\xrightarrow{w*} Q$ we obtain that  $\min_{P\in \P,Q\in \Q}\int_{x}f\,dP - \int_{x}f\,dQ\leq  \min_{P\in\cco \P,Q\in\cco \Q}\left(\int_{x}f\,dP - \int_{x}f\,dQ\right)$.


\subsection{Proof of Corollary \ref{corollary: approximate Gordan}}

If system I has $\bar f$ as a solution, then for all $(P,Q)\in \P\times \Q $ we have
\begin{align*}
d_{1}(P,Q) &= \sup_{f\in C(X),\norm{f}_{\infty}\leq 1}\left(\int_{X}f\,dP -\int_{X}f\,dQ\right)\geq \int_{X}\bar f\,dP -\int_{X}\bar f\,dQ >\epsilon,
\end{align*}
so system II has none. In view of Theorem \ref{theorem: duality for probabilities}, when system I has no solution, if we choose $\bar P\in\cco\P$ and  $\bar Q\in\cco\Q$ that minimize the distance $d_{1}(P,Q)$ then we must have $d_{1}(\bar P,\bar Q)\leq \epsilon$.

\subsection{Proof of Corollary \ref{corollary: version of original epsilon alternative}}
Because of Corollary \ref{corollary: approximate Gordan}, the existence of probability measures $P\in\cco\P$ and $Q\in\cco\Q$ with $\ell_{1}$ distance at most $\epsilon$ is equivalent to system I in Corollary \ref{corollary: approximate Gordan} having no solution. This means that, after employing standard arguments involving continuity and compactness in the weak$^{\ast}$ topology, the inequality in (\ref{equation: inequality with epsilon separate}) holds for any continuous function $f$ of norm one.


\subsection{Proof of Proposition \ref{proposition: characterize convex combination residual}}
Suppose that for some $(P,Q)\in\cco \P\times\cco \Q$, and $R\in\R$, the expression in (\ref{equation: expression for general residual behavior}) holds. Let $f\in C(X)$ be such that $\norm{f}_{\infty}=1$.  Apply a similar reasoning to that leading to (\ref{equation: equality in sup min}) in the proof of Theorem \ref{theorem: duality for probabilities} to obtain
\begin{align}
	\min_{\Ph\in\cco \P}\int_{X}f\,d\Ph &= \min_{\Ph\in \P}\int_{X}f\,d\Ph\notag \\
	&\leq \int_{X}f\,dP\notag\\
	& = (1-\epsilon) \int_{X}f\,dQ + \epsilon\int_{X}f\,dR. \label{equation: expression to be bounded}
\end{align}
The condition in (\ref{equation: condition for general residual behavior}) now becomes a consequence of bounding the terms in the expression in (\ref{equation: expression to be bounded}): $ \int_{X}f\,dQ \leq \max_{\widehat Q\in\cco \Q}\int_{X}f\,d\widehat Q$, and $\int_{X}f\,dR\leq  \max_{\widehat R\in\R}\int_{X}f\,d\widehat R$.

Define $\widehat \Q = \{(1-\epsilon) Q+\epsilon R: Q\in \cco \Q, R\in\R\}$, which is compact and convex. If $\cco \P\cap \widehat \Q = \emptyset$, by the Strong Separating Hyperplane Theorem (Theorem 5.79 in Aliprantis and Border, 2006) there exists a nonzero function $f\in C(X)$ such that
\begin{align*}
	\min_{ P\in\cco \P}\int_{X}f\,d P&> \max_{\widehat Q\in\widehat\Q}\int_{X}f\,d\widehat Q =(1-\epsilon)\max_{Q\in\cco \Q}\int_{X}f\,dQ+\epsilon \max_{R\in\R}\int_{X}f\,dR.
\end{align*}
Let $\sigma = \min_{x\in X} f(x)$ and $g = f - \sigma \indic_{X}\geq 0$. Note that $\norm{g}_{\infty} = \max_{x\in X} g(x)$. Combining this observation with the last inequality, we obtain that
\begin{align*}
	\min_{ P\in\cco \P}\int_{X}\widehat g\,d P> (1-\epsilon)\max_{Q\in\cco \Q}\int_{X}\widehat g\,dQ+\epsilon \max_{R\in\R}\int_{X}\widehat g\,dR,
\end{align*} 
upon defining $\widehat g = \frac{g}{\norm{g}_{\infty}}$, which has norm one.


\subsection{Proof of Proposition \ref{proposition: aggregation and max min condition}}

The following fact proves to be useful.

\begin{fact}\label{fact: useful}
	For all $m\in \Delta(\Q)$ and $f\in C(X)$,
	\begin{align}
		\int_{X}f\, d Q_{m} = \int_{\Q}\left( \int_{X}f\, dQ\right)dm.\label{equation: equality averaging average}
	\end{align}
	\begin{proof}
		First let $\Delta_{s}(\Q)$ denote the set of simple probability measures on $\Q$ (the relevant definition is given in the proof of Proposition \ref{proposition: weighted average of probabilities} above). For $m\in\Delta_{s}(\Q)$ the equality $\int_{X}f\, d Q_{m} = \int_{\Q}\left( \int_{X}f\, dQ\right)dm$ is  a consequence of the linearity of the linear functional induced on the space of signed measures of bounded variation on $X$. In fact, in this case $\int_{X}f\, d Q_{m}  = \sum_{i=1}^{k}\lambda_{i}\left(\int_{X}f\,dQ_{i}\right)$ when $m = \sum_{i=1}^{k}\lambda_{i}\delta_{Q_{i}}$. For a general $m\in\Delta(\Q)$, note that by the Density Theorem (Theorem 15.10 in Aliprantis and Border, 2006) we also know that $\Delta(\Q)=  \cl_{w*} \Delta_{s}(\Q) $, so that $m_{n}\xrightarrow{w*} m $ for some sequence $(m_{n})$ in $\Delta_{s}(\Q)$. The mapping $Q\mapsto \int_{X}f\,dQ$ is weak* continuous (Theorem 15.5 in Aliprantis and Border, 2006). By weak convergence we therefore know that $\lim_{n\to\infty} \int_{\Q}\left( \int_{X}f\, dQ\right)dm_{n} =  \int_{\Q}\left( \int_{X}f\, dQ\right)dm.$  At the same time, since for any closed subset $E$ of $X$ the indicator function $\indic_{E}$ on $X$ is upper semicontinuous, the function $Q\mapsto Q(E) = \int_{X}\indic_{E}\,dQ$ is also upper semicontinuous, and so is the mapping $m\mapsto Q_{m}(E)$ (see Theorem 15.5 in Aliprantis and Border, 2006).  Therefore, for any closed set $E\subseteq X$ we have $ \limsup_{n}Q_{m_{n}}(E)\leq Q_{m}(E).$ 	By the characterization of weak convergence we know that $Q_{m_{n}}\xrightarrow{w*} Q_{m}$. Combining the convergence of $(Q_{m_{n}})$ and $(m_{n})$ just established yields
		 \begin{align*}
		 	\int_{X}f\, d Q_{m} = \lim_{n\to\infty}	\int_{X}f\, d Q_{m_{n}} =  \lim_{n\to\infty} \int_{\Q}\left( \int_{X}f\, dQ\right)dm_{n} =  \int_{\Q}\left( \int_{X}f\, dQ\right)dm.&\qedhere
		 \end{align*}
		\end{proof}
	\end{fact}
		
Turn to the proof of Proposition \ref{proposition: aggregation and max min condition}. Note that for any $f\in C(X)$ with unit norm the function $-f$ also has norm one, and 
\begin{align*}
	\int_{X}(-f)\,dP\leq \max_{Q\in\Q}\int (-f)\,dQ + \epsilon\Leftrightarrow \int_{X}f\,dP\geq \min_{Q\in\Q}\int f\,dQ - \epsilon.
\end{align*} 
Therefore the statements in (i) and (ii) are equivalent. Now assume that $P$ is as in (\ref{equation: expression for P}). In view of equation (\ref{equation: equality averaging average}), simple calculations reveal that 
\begin{align}
	\int_{X}f\,dP &= \int_{\Q}\left( \int_{X}f\, dQ\right)dm + \int_{X}f\,de\notag \\
				 & \leq  	\max_{Q\in\Q}\int_{X}f\,dQ + \int_{X}f\,de\notag \\
				 &\leq 	\max_{Q\in\Q}\int_{X}f\,dQ + \norm{f}_{\infty}\norm{e}_{1}.\label{equation: last estimate inequality for aggregation}
				 \end{align}
For $\norm{f}_{\infty}=1$ and $\norm{e}_{1}\leq \epsilon$ the inequality in item (i) is now a consequence of (\ref{equation: last estimate inequality for aggregation}). Conversely, suppose that $P$ cannot be expressed as in (\ref{equation: expression for P}). By Corollary \ref{corollary: approximate Gordan} there exists a continuous function $f\colon X\to\Re$ with unit norm such that $\int_{X}f\,dP -\int_{X}f\,dQ  >\epsilon$ for all $Q\in \Q$. In particular, because of compactness of $\Q$ and continuity of the mapping $Q\mapsto \int_{X}f\,dQ$ we have that $\int_{X}f\,dP>\max_{Q\in\Q}\int_{X}f\,dQ + \epsilon$, thus showing that the statement in item (i) does not hold. This establishes the converse implication.


\subsection{Proof of Proposition \ref{proposition: aggregation with version of Pareto}}

Let $P$ be as in (\ref{equation: expression for P}). Suppose that $f,g\in C(X)$ are such that for all $Q\in \Q$ we have that $\int_{X}f\,dQ\geq \int_{X}g\,dQ$. Then $\int_{\Q}\left(\int_{X}f\,dQ\right)dm\geq \int_{\Q}\left(\int_{X}g\,dQ\right)dm$ in view of the monotonicity of the integral with respect to the probability measure $m\in\Delta(\Q)$. Let $e_{+},e_{-}\in \ca(X)$ denote nonnegative measures such that $e = e_{+}-e_{-}$ (their existence is assured by the Hahn decomposition theorem). Note that $\norm{e}_{1} = \norm{e_{+}}_{1} +\norm{e_{-}}_{1}$ and that $0 = e(X) = \norm{e_{+}}_{1} -\norm{e_{-}}_{1} $. Hence
\begin{align*}
	\int_{X}f\,dP &= \int_{X}f\,dQ_{m} + \int_{X}f\,de_{+}-\int_{X}f\,de_{-}\\
	&\geq \int_{X}g\,dQ_{m} + \int_{X}f\,de_{+}-\int_{X}f\,de_{-}\\
	& = \int_{X}g\,dP+ \int_{X}(f-g)\,de_{+}-\int_{X}(f-g)\,de_{-}\\
	&\geq  \int_{X}g\,dP + \norm{e_{+}}_{1} \min_{x\in X}[f(x)-g(x)]- \norm{e_{-}}_{1}\max_{x\in X}[f(x)-g(x)]\\
	& =  \int_{X}g\,dP  - \frac{\norm{e}_{1}\omega(f-g)}{2} \\
	&\geq \int_{X}g\,dP  - \frac{\epsilon\cdot\omega(f-g)}{2}.
\end{align*}

Now assume that $P$ cannot be expressed as in (\ref{equation: expression for P}). Then system II in Corollary \ref{corollary: approximate Gordan} has no solution. Consequently there exists $f\in C(X)$ with $\norm{f}_{\infty}=1$ and $\int_{X}f\,dP<\int_{X}f\,dQ-\epsilon$ for all $Q\in\Q$. This is equivalent to saying that, for $g= \left(\min_{Q\in\Q}\int_{X}f\,dQ\right)\cdot \indic_{X}$, $\int_{X}f\,dP<\int_{X}g\,dP-\epsilon$. Since $\omega(f)  = \omega(f-g)$ and $\omega(f)\leq 2\norm{f}_{\infty}$, we obtain that $\int_{X}f\,dP<\int_{X}g\,dP-\frac{\epsilon\cdot\omega(f-g)}{2}$. By construction $\int_{X}f\,dQ\geq \int_{X}g\,dQ$ for all $Q\in\Q$. This gives a violation of condition $\CC_{\epsilon}$.

 
\subsection{Proof of Proposition \ref{proposition: a la Genest}}

Suppose that $P$ can be expressed as in (\ref{equation: aggregation a la Genest}). Let $f,g\in C(X)$ be such that $\int_{X}f\,dQ\geq \int_{X}g\,dQ$ for all $Q\in\Q$. By Fact \ref{fact: useful} above and nonnegativity of the mapping $Q\mapsto \int_{X}(f-g)\,dQ$, we have $\int_{X}f\,dQ_{m}\geq\int_{X}g\,dQ_{m}$. Hence
\begin{align*}
	\int_{X}f\,dP& = (1-\epsilon) \int_{X}f\,dQ_{m} + \epsilon \int_{X}f\,dR\\
					& \geq (1-\epsilon) \int_{X}g\,dQ_{m} + \epsilon \int_{X}f\,dR\\
					& = \int_{X}g\,dP + \epsilon \int_{X}(f-g)\,dR\\
					& \geq  \int_{X}g\,dP +\epsilon\min_{x\in X}[f(x)-g(x)]\\
					& =  \int_{X}g\,dP- \epsilon\omega(f-g) +\epsilon \max_{x\in X}[f(x)-g(x)].
\end{align*}

Now assume that $P$ cannot be expressed as in (\ref{equation: aggregation a la Genest}). It follows from Proposition \ref{proposition: characterize convex combination residual} in the case $\P = \{P\}$ and $\R=\Delta(X)$ that for some $f\in C(X)$ we have 
\begin{align}\label{equation: inequality sufficiency part residual Pareto }
	\int_{X}f\,dP< (1-\epsilon)\min_{Q\in\Q}\int_{X}f\,dQ + \epsilon \min_{x\in X}f(x).
\end{align}
By defining $g$ as the constant function $\left(\min_{Q\in\Q}\int_{X}f\,dQ\right)\cdot \indic_{X}$ we have by construction $\int_{X}f\,dQ\geq \int_{X}g\,dQ$ for al $Q\in\Q$. At the same time, the inequality in (\ref{equation: inequality sufficiency part residual Pareto }) gives
\begin{align*}
	\int_{X}f\,dP< \int_{X}g\,dP  +\epsilon \min_{x\in X}[f(x) - g(x)],
\end{align*}
thus contradicting condition $\CC_{\epsilon}^{\ast}$.


\subsection{Proof of Proposition \ref{proposition: aggregation with version of Pareto nonatomic}}

Suppose that $P = \sum_{i=1}^{N}m_{i}Q_{i}+e$, with $m_{i}\geq 0$ ($i=1,\dots,N)$, $e\in \ca(X)$, and $\norm{e}_{1}\leq \epsilon$. Let $E_{1},E_{2}\in\Sigma_{X}$ be such that $Q_{i}(E_{1})\geq Q_{i}(E_{2})$ for all $i=1,\dots,N$. Using the Hahn decomposition theorem we know that there are sets $X_{-},X_{+}\in\Sigma_{X}$ such that $\norm{e}_{1} = e(X_{+}) -e(X_{-})  $. By the construction of the sets $X_{+}$ and $X_{-}$ in the mentioned theorem we know that $e(E_{1})\geq e(X_{-})$ and $e(E_{2})\leq e(X_{+})$. Then
\begin{align*}
	P(E_{1}) &= \sum_{i=1}^{N}m_{i}Q_{i}(E_{1}) +e(E_{1})\\
		& \geq P(E_{2}) + e(E_{1}) - e(E_{2}) \\
		& \geq P(E_{2}) + e(X_{-}) - e(X_{+})\\
		& = P(E_{2}) -\norm{e}_{1}\\
		&\geq P(E_{2}) -\epsilon.
	\end{align*}
	
Now assume that $P$ and $\Q$ satisfy condition $\CC^{M}_{\epsilon}$. This part of the proof  follows similar steps to those of Proposition 2 in Mongin (1995). Define the set $C\subseteq\Re^{N+1}$ so that
\begin{align*}
	C = \{z : z_{i} = Q_{i}(E_{1})-Q_{i}(E_{2}) \text{ for }1\leq i\leq N, z_{N+1} = P(E_{1})-P(E_{2}), E_{1},E_{2}\in\Sigma_{X} \}.
\end{align*}
Obviously $C\neq \emptyset$. By the Lyapunov Convexity Theorem (Theorem 13.33 in Aliprantis and Border, 2006) the set $C$ is convex and compact. We also define the set $D \subseteq \Re^{N+1}$ so that
\begin{align*}
	D = \{z: z_{i}\geq 0\text{ for }1\leq i\leq N, z_{N+1}<-\epsilon\}.
\end{align*}
The set $D$ is also nonempty and convex. Condition $\CC^{M}_{\epsilon}$ implies that $C\cap D = \emptyset$. By the  Separating Hyperplane Theorem (e.g., Theorem 7.30 in Aliprantis and Border, 2006) there exists a nonzero vector $\lambda\in\Re^{N+1}$ such that 
\begin{align*}
	\sum_{i=1}^{N}\lambda_{i}[Q_{i}(E_{1})-Q_{i}(E_{2})] + \lambda_{N+1}[P(E_{1})-P(E_{2})]\geq \sum_{i=1}^{N}\lambda_{i}z_{i}+\lambda_{N+1}z_{N+1}
\end{align*}
for all $E_{1},E_{2}\in\Sigma_{X}$, and $z\in D$. Setting $E_{1}=E_{2}$,  $z_{i} = 0$ for all $i=1,\dots,N$, and $z_{N+1}=-2\epsilon$, we have $0\geq -2\epsilon \lambda_{N+1}$, and thus $\lambda_{N+1}\geq 0$. If $\lambda_{N+1} = 0$, since $(1,\dots,1,-2\epsilon)$ belongs to the interior of $D$, we have $0>\sum_{i=1}^{N}\lambda_{i}$. At the same time, by setting $E_{1} = X$, $E_{2}=\emptyset$, and $z$ equal to the zero vector, we obtain that $\sum_{i=1}^{N}\lambda_{i}\geq 0$, a contradiction. Therefore, we may assume without loss of generality that $\lambda_{N+1}=1$. Also note that for $E_{1}=E_{2}$, $z_{N+1} =-2\epsilon$, $z_{i} = n$, and $z_{j}=0$ for $j\neq i$ we have that $0\geq -2\epsilon  +\lambda_{i}n\Leftrightarrow \lambda_{i}\leq \frac{2\epsilon}{n}$. Taking the limit in the last inequality as $n\to\infty$ yields $\lambda_{i}\leq 0$. Define $m_{i} = -\lambda_{i}$, for all $i=1,\dots,N$. For any two sets $E_{1},E_{2}\in\Sigma_{X}$, and $z_{i}=0$ for all $i=1,\dots,N$, upon letting $z_{N+1}\uparrow-\epsilon$ we have that
\begin{align*}
	P(E_{1})-P(E_{2}) - \sum_{i=1}^{N}m_{i}[Q_{i}(E_{1}) -Q_{i}(E_{2}) ]\geq -\epsilon.
\end{align*}
By exchanging the roles of $E_{1}$ and $E_{2}$ it follows that, for $e= P - \sum_{i=1}^{N}m_{i}Q_{i}\in\ca(X)$,
\begin{align}\label{equation: inequality for e proof}
	\abs{e(E_{1})-e(E_{2})}\leq \epsilon.
\end{align}
By the Hahn decomposition theorem, for $X_{+},X_{-}\in\Sigma_{X}$ such that $\norm{e}_{1} = e(X_{+}) - e(X_{-})$, we obtain that $\norm{e}_{1}\leq \epsilon$ by setting $E_{1}=X_{+}$  and $E_{2} = X_{-}$  in (\ref{equation: inequality for e proof}).


\subsection{Proof of Proposition \ref{proposition: aggregation with version of Pareto nonatomic 2}}

Assume that $P= \sum_{i=1}^{N}m_{i}Q_{i}+e$ with $m$ in the $N-1$ dimensional simplex and $\norm{e}_{1}\leq \epsilon$. Equivalence of both conditions follows from taking complements. So we only show that $P$ satisfies (ii). Using the Hahn decomposition theorem we know that there are disjoint sets $X_{-},X_{+}\in\Sigma_{X}$ such that $\norm{e}_{1} = e(X_{+}) -e(X_{-})  $. In the decomposition we have $e(X_{-}) = \inf_{E\in\Sigma_{X}}e(E)$. At the same time, $e(X)= 0$, so that $e(X_{+}) =-e(X_{-})  $ and thus $ e(X_{-})  \geq -\frac{\epsilon}{2} $. Then
\begin{align*}
	P(E)& = \sum_{i=1}^{N}m_{i}Q_{i}(E) + e(E)\\
		&\geq \min_{i}Q_{i}(E) +e(X_{-})\\
		&\geq \min_{i}Q_{i}(E) -\frac{\epsilon}{2}. 
\end{align*}

Now assume that condition (ii) holds. Let $f\in C(X)$ be non-constant and such that $\norm{f}_{\infty}=1$. Let $\sigma=  \min_{x\in X}f(x)$, and $g = \frac{f-\sigma\indic_{X}}{\norm{f-\sigma\indic_{X}}}$. Clearly $g$ is continuous and $0\leq g\leq 1$. By the Lyapunov Convexity Theorem (Theorem 13.33 in Aliprantis and Border, 2006), there exists $E\in\Sigma_{X}$ such that $P(E) = \int_{X}g\,dP$, and $Q_{i}(E)=\int_{X}g\,dQ_{i}$, for $i=1,\dots,N$. From (ii), $P(E)\geq \min_{i}Q_{i}(E)-\frac{\epsilon}{2} $. Since $\norm{f-\sigma\indic_{X}}_{\infty}\leq 2$, and thus $-\frac{1}{2}\geq -\frac{1}{\norm{f-\sigma\indic_{X}}_{\infty}}$, we get
\begin{align*}
	P(E)\geq \min_{i}Q_{i}(E) -\frac{\epsilon}{\norm{f-\sigma\indic_{X}}_{\infty}},
\end{align*}
which is equivalent to
\begin{align}\label{equation: last inequality mongin 2}
	 \int_{X}f\,dP\geq \min_{i} \int_{X}f\,dQ_{i} -\epsilon
\end{align}
after simple algebra. The inequality in (\ref{equation: last inequality mongin 2}) is condition (ii) in Proposition \ref{proposition: aggregation and max min condition}, and this completes the proof.


\subsection{Proof of Proposition \ref{proposition: first epsilon random RUM}}

Suppose that $P_{0}$ is given as in (\ref{equation: RUM plus e}). Let $e_{+}$ and $e_{-}$, both in $\Re^{M}$, denote the positive and the negative parts of $e$, respectively. Put $\Nh = 2^{\Ny}-1$. Since
\begin{align*}
	\Nh &= \sum_{(y,Y)\in X}P_{0}(y,Y)\\
		 &= \sum_{\spref\in\L}\sum_{(y,Y)\in X}a_{\spref}(y,Y)\pi(\spref) + \norm{e_{+}}_{1} - \norm{e_{-}}_{1},
\end{align*}
and using the fact that, given $\spref$, for each nonempty set $Y\subseteq\Y$ there is exactly one $y$ for which $a_{\spref}(y,Y)=1$, we obtain 
\begin{align*}
	\Nh = \sum_{\spref\in\L}\Nh\pi(\spref) + \norm{e_{+}}_{1} - \norm{e_{-}}_{1},
\end{align*}
and thus $\norm{e_{+}}_{1}= \norm{e_{-}}_{1} = \frac{\norm{e}_{1}}{2}\leq \frac{\epsilon}{2}$. Let $T$ be a trial sequence of width $w_{T}$. Then
\begin{align*}
	\sum_{i=1}^{M}P_{0}(y_{i},Y_{i})t_{i} &= \sum_{\spref\in\L}\pi(\spref)\left[ \sum_{i=1}^{M}a_{\spref}(y_{i},Y_{i}) t_{i}\right]+ \sum_{i=1}^{M}e_{+}(i)t_{i}-\sum_{i=1}^{M}e_{-}(i)t_{i}
\end{align*}
In view of the inequalities
\begin{align*}
	 \sum_{\spref\in\L}\pi(\spref)\left[ \sum_{i=1}^{M}a_{\spref}(y_{i},Y_{i}) t_{i}\right]\leq \max_{\spref\in\L}\left[ \sum_{i=1}^{M}a_{\spref}(y_{i},Y_{i}) t_{i}\right]
\end{align*}
and
\begin{align*}
	\sum_{i=1}^{M}e_{+}(i)t_{i}-\sum_{i=1}^{M}e_{-}(i)t_{i}&\leq \norm{e_{+}}_{1}\max_{i}t_{i} -  \norm{e_{-}}_{1}\min_{i}t_{i}\\
	& \leq \frac{\epsilon}{2}(\max_{i}t_{i} -\min_{i}t_{i})\\
	&=\frac{w_{T}\epsilon}{2}
\end{align*}
the stochastic choice function satisfies $\epsilon$-ARSP.

Now assume that $\epsilon$-ARSP holds. With respect to the vectors $P$ and $Q_{\spref}$ mentioned in the text this is equivalent to saying that, for every trial sequence $T$ of width $w_{T}$,
\begin{align}
	\sum_{i=1}^{M}P(y_{i},Y_{i})t_{i} \leq \max_{\spref\in\L}\sum_{i=1}^{M}Q_{\spref}(y_{i},Y_{i})t_{i} + \frac{w_{T}\epsilon}{2\Nh}.\label{equation: implication epsilon ARSP P and Q}
\end{align}
To show that $P_{0}$ can be represented as in equation (\ref{equation: RUM plus e}), it suffices to show that the set $\P = \{P\}$ and the convex hull of $\Q =  \{Q_{\spref}:\,\spref\,\in\L\}$ have points whose distance is at most $ \frac{\epsilon}{\Nh}$. Let $\bar\epsilon=\frac{\epsilon}{\Nh}$. Because of Corollary \ref{corollary: approximate Gordan}, it suffices to show that for no vector $z\in\Re^{M}$ with $\norm{z}_{\infty}=1$ we have
\begin{align}
	\sum_{i=1}^{M}P_{i}z_{i} > \sum_{i=1}^{M}Q_{i}z_{i} + \bar\epsilon \quad \text{ for all }Q\in\Q,\label{equation: inequality strict with z}
\end{align}
where $P_{i}$ and $Q_{i}$ are the $i$-th components of $P$ and $Q$, respectively.

So suppose that such a vector exists, and denote it by $z^{\ast}$. We will show that this leads to a contradiction. By continuity of the linear functionals on Euclidean spaces of dimension $M$, for each $Q\in \Q$ there exists $\gamma_{Q}>0$ such that the inequality in (\ref{equation: inequality strict with z}) holds for all $z\in\Re^{M}$ with $\norm{z-z^{\ast}}_{\infty}< \gamma_{Q}$. By letting $\gamma$ denote the minimum among the $\gamma_{Q}$, for each coordinate $i$ such that $-1<z^{\ast}_{i}<1$, choose a rational number $z_{i}$ sufficiently close to $z^{\ast}_{i}$. For the coordinates $i$ such that $z_{i}^{\ast}=\pm 1$, put $z_{i} = z_{i}^{\ast}$. By the denseness of $\mathbb Q$ in $\Re$, we can therefore choose points so that the vector $z$ has rational entries, and moreover $\norm{z-z^{\ast}}_{\infty}<\gamma$. Consequently, it satisfies (\ref{equation: inequality strict with z}) as well and furthermore $\norm{z}_{\infty}=1$. Now define $\zt\in \mathbb Q^{M}$ so that $\zt_{i} = z_{i}-\min_{j}z_{j}\geq 0$. By construction $\zt_{i} = \frac{p_{i}}{q_{i}}$ for some $0\leq p_{i}\in\mathbb Z$ and $0< q_{i}\in\mathbb Z$, for all $i=1,\dots,M$. Note that $\zt$ also satisfies the inequality in (\ref{equation: inequality strict with z}), which becomes, after some algebra,
\begin{align}
	\sum_{i=1}^{M}P_{i}t_{i} > \max_{Q\in\Q}\sum_{i=1}^{M}Q_{i}t_{i} +\bar\epsilon\prod_{j}q_{j},\label{equation: expression with ratios strict inequality}
\end{align}
where $t_{i} = p_{i}\prod_{j\neq i}q_{j}$. Note that for at least one $i$ we have $p_{i}=0$. At the same time, since $\omega(\zt) =\omega(z)\leq 2\norm{z}_{\infty} =2$, we know that, for all $i$,
\begin{align}
	p_{i}\prod_{j\neq i}q_{j}\leq 2 \prod_{j}q_{j}.\label{equation: inequality product}
\end{align}
Combining the inequalities in (\ref{equation: expression with ratios strict inequality}) and (\ref{equation: inequality product}) we obtain that
\begin{align*}
	\sum_{i=1}^{M}P_{i}t_{i}>  \max_{Q\in\Q}\sum_{i=1}^{M}Q_{i}t_{i} + \frac{\max_{i}t_{i}}{2}\bar\epsilon.
\end{align*}
Since $\min_{i}t_{i}=0$, we know that $w_{T} = \max_{i}t_{i}$ represents the width of the tagged trial sequence constructed from the integers $t_{i}$. This is a contradiction with (\ref{equation: implication epsilon ARSP P and Q}).


\subsection{Proof of Proposition \ref{proposition: second epsilon random RUM}}

Assume that $P_{0}$ is expressed as in (\ref{equation: nearly RUM residual behavior}). If $T$ is a tagged trial sequence, simple calculations show that
\begin{align*}
	\sum_{i=1}^{M}P_{0}(y_{i},Y_{i})t_{i} &= (1-\epsilon)\sum_{\spref\in\L}\pi(\spref) \left[\sum_{i=1}^{M}a_{\spref}(y_{i},Y_{i}) t_{i}\right]+\epsilon\sum_{i=1}^{M}R_{0}(y_{i},Y_{i})t_{i}\\
								&\leq (1-\epsilon)\max_{\spref\in\L}\sum_{i=1}^{M}a_{\spref}(y_{i},Y_{i})t_{i} + \epsilon(2^{\Ny}-1)\max_{i}t_{i}.
\end{align*}
This is $\epsilon$-ARSP$^{\ast}$.

Conversely, suppose that  $\epsilon$-ARSP$^{\ast}$ holds.  Put $\Nh = 2^{\Ny}-1$. Because of Proposition  \ref{proposition: characterize convex combination residual} applied to $\P=\{P\}$, $\Q = \{Q_{\spref}:\spref\,\in\L\}$, and $\R =\{\frac{1}{\Nh} R\in\Re^{M}: R_{0}(y,Y)\geq 0,  \sum_{y\in Y}R(y,Y)=1\text{ for all } Y\subseteq \Y, Y\neq \emptyset \}$, it suffices to show that for no vector $z\in\Re^{M}$ with nonnegative components and $\norm{z}_{\infty}=1$ we have 
\begin{align}\label{equation: strict inequality no residual behavior}
	\sum_{i=1}^{M}P_{i}z_{i}&>(1-\epsilon)\max_{Q\in\Q}\sum_{i=1}^{M}Q_{i}z_{i} + \epsilon\max_{R\in\R}\sum_{i=1}^{M}R_{i}z_{i}.
\end{align}
So suppose by way of contradiction that the vector $z^{\ast}\in\Re^{M}$ with nonnegative components verifies the inequality in (\ref{equation: strict inequality no residual behavior}).  This is equivalent to
\begin{align}\label{equation: strict inequality no residual behavior 2}
	\sum_{i=1}^{M}P_{0}(y_{i},Y_{i})z_{i}^{\ast}>(1-\epsilon)\max_{\spref\in\L}\sum_{i=1}^{M}a_{\spref}(y_{i},Y_{i})z_{i}^{\ast} +\epsilon \max_{R_{0}\in\R_{0}}\sum_{i=1}^{M}R_{0}(y_{i},Y_{i})z_{i}^{\ast},
\end{align} 
where $\R_{0} = \{\Nh R: R\in \R\}$. For a fixed $Y$ the mappings $y\mapsto P_{0}(y,Y)$, $y\mapsto a_{\spref}(y,Y)$ and $y\mapsto R_{0}(y,Y)$ are probability measures on $Y$. We now identify $z(y,Y)$ with the corresponding coordinate of $z$ for which  $(y_{i},Y_{i}) = (y,Y)$. We can now add $1-\max_{y\in Y}z^{\ast}(y,Y)$ to both sides of (\ref{equation: strict inequality no residual behavior 2}) and adjust the vector accordingly so that such an inequality also holds when $\max_{y\in Y}z^{\ast}(y,Y)=1$. By repeating this procedure for each nonempty subset $Y$ of $\Y$, we can assume without any loss of generality that for all $Y\subseteq \Y$ with $Y\neq\emptyset$ we have  $z^{\ast}(y,Y)=1$ for some $y\in Y$. Now proceed as in the proof of Proposition \ref{proposition: first epsilon random RUM} to adjust again the vector $z^{\ast}$ if needed so that $z^{\ast}\in \mathbb Q^{M}$, $z_{i}^{\ast}\geq 0$ for all $i=1,\dots,M$, and the coordinates for which $z_{i}^{\ast}=1$ are preserved. Hence we can write $z^{\ast}_{i} =\frac{p_{i}}{q_{i}}$ with $0\leq p_{i}\in\mathbb Z$, and $0< q_{i}\in \mathbb Z$. Let $t_{i} = p_{i}\prod_{j\neq i}q_{j}$. We therefore have
\begin{align*}
	\sum_{i=1}^{M}P_{0}(y_{i},Y_{i})t_{i}&>(1-\epsilon)\max_{\spref\in\L}\sum_{i=1}^{M}a_{\spref}(y_{i},Y_{i})t_{i} +\epsilon \max_{R_{0}\in\R_{0}}\sum_{i=1}^{M}R_{0}(y_{i},Y_{i})t_{i}^{\ast}\\
	& = (1-\epsilon)\max_{\spref\in\L}\sum_{i=1}^{M}a_{\spref}(y_{i},Y_{i})t_{i} +\epsilon (2^{\Ny}-1)\max t_{i}^{\ast},
\end{align*}
where the last equality follows from maximizing the second term in the right-hand side of the inequality in (\ref{equation: strict inequality no residual behavior 2}). This contradicts $\epsilon$-ARSP$^{\ast}$.


\section*{References}

\noindent Acciaio, B., Backhoff, J.,  Gudmund, P. (2022). Quantitative Fundamental Theorem of Asset Pricing. \textit{arXiv preprint arXiv:2209.15037}.

\medskip

\noindent Aliprantis, C.\ D., Border, K.\ C.\ (2006). \emph{Infinite Dimensional Analysis: a Hitchhiker's Guide}, 3rd ed., New York: Springer.

\medskip 

\noindent Apesteguia, J., Ballester, M. A. (2021). Separating predicted randomness from residual behavior.  \textit{Journal of the European Economic Association}, 19, 1041-1076.

\medskip

\noindent Araujo, A.\, Chateauneuf, A., Faro, J.\ H. (2012). Pricing rules and Arrow–Debreu ambiguous valuation. \textit{Economic Theory}, 49, 1-35.

\medskip

\noindent Araujo, A.\, Chateauneuf, A., Faro, J.\ H. (2018). Financial market structures revealed by pricing rules: efficient complete markets are prevalent. \textit{Journal of Economic Theory}, 173, 257-288.

\medskip

\noindent Beggs, A. (2021). Afriat and arbitrage.  \textit{Economic Theory Bulletin}, 9, 167-176.

\medskip

\noindent Block H. D., Marschak, J. (1960). ``Random orderings and stochastic theories of responses,'' in I. Olkin, et al. (eds.), \textit{Contributions to Probability and Statistics: Essays in Honor of Harold Hotelling},
 Stanford University Press, 97-132.  

\medskip

\noindent Bogachev, V.\ I. (2007). \emph{Measure Theory}, New York: Springer.

\medskip

\noindent Bogachev, V.\ I. (2018). \emph{Weak Convergence of Measures}, Providence: American Mathematical Society.

\medskip

\noindent Border, K.\ C.\ (2007). \emph{Introductory Notes on Stochastic Rationality.} California Institute of Technology. 

\medskip

\noindent Chateauneuf, A.\, Cornet, B.\ (2022). Submodular financial markets with frictions. \textit{Economic Theory}, 73, 721-744.
\medskip

\noindent Clark, S.\ A.\ (1996). The random utility model with an infinite choice space. \textit{Economic Theory}, 7, 179-189.
\medskip

\noindent Dax, A. (2006). The distance between two convex sets. \textit{Linear Algebra and its Applications}, 416, 184-213.

\medskip

\noindent Dax, A., Sreedharan, V.\ P. (1997). Theorems of the alternative and duality. \textit{Journal of Optimization Theory and Applications}, 94, 561-590.

\medskip

\noindent Falmagne, J.\ C.\ (1978). A representation theorem for finite random scale systems. \textit{Journal of Mathematical Psychology}, 18, 52-72.

\medskip

\noindent Gibbs, A.\ L., Su, F.\ E. (2002). On choosing and bounding probability metrics. \textit{International Statistical Review}, 70, 419-435.

\medskip

\noindent Genest, C.\ (1984). Pooling operators with the marginalization property. \textit{Canadian Journal of Statistics}, 12, 153-163.

\medskip

\noindent G{\"u}ler, O., Hoffman, A.\ J., Rothblum, U.\ G.\ (1995). Approximations to solutions to systems of linear inequalities. \textit{SIAM Journal on Matrix Analysis and Applications}, 16, 688-696.

\medskip

\noindent Hellman, Z.\ (2013). Almost common priors. \textit{International Journal of Game Theory}, 42, 399-410.

\medskip

\noindent Hellman, Z., Pint\'{e}r, M. (2022). Charges and bets: a general characterisation of common priors. \textit{International Journal of Game Theory}, 51, 567-587.

\medskip

\noindent Hoffman, A.\ J.\ (1952). On approximate solutions of systems of linear inequalities. \textit{Journal of Research of the National Bureau of Standards}, 49, 263-265.

\medskip

\noindent Huber, P.\ J.\ (1964). Robust estimation of a location parameter. \textit{The Annals of Mathematical Statistics}, 73-101.

\medskip

\noindent Kneser, H. (1952). Sur un th\'{e}oreme fondamental de la th\'{e}orie des jeux. \textit{C.\ R.\ Acad.\ Sci.\ Paris}, 234, 2418-2420.

\medskip

\noindent Luenberger, D.\ G. (1997). \emph{Optimization by Vector Space Methods}, New York: John Wiley \& Sons.

\medskip

\noindent McConway, K.\ J.\ (1981). Marginalization and linear opinion pools. \textit{Journal of the American Statistical Association}, 76, 410-414.

\medskip

\noindent McFadden, D. (2005). Revealed stochastic preference: a synthesis. \textit{Economic Theory}, 26, 245-264.

\medskip

\noindent McFadden, D., Richter, K.\ (1990) ``Stochastic Rationality and Revealed Stochastic Preference,'' in J. Chipman, D. McFadden, K. Richter (eds.), \textit{Preferences, Uncertainty, and Optimality}, Westview Press, 161-186.

\medskip

\noindent Mongin, P.\ (1995). Consistent Bayesian aggregation. \textit{Journal of Economic Theory}, 66, 313-351.

\medskip

\noindent Nau, R.\ F.\ (1992). Joint coherence in games of incomplete information. \textit{Management Science}, 38, 374-387.

\medskip

\noindent Nau, R.\ F.\ (1995a). The incoherence of agreeing to disagree. \textit{Theory and Decision}, 39, 219-239.

\medskip

\noindent Nau, R.\ F.\ (1995b). Coherent decision analysis with inseparable probabilities and utilities. \textit{Journal of risk and uncertainty}, 10, 71-91.
\medskip

\noindent Nau, R. (2015). Risk-neutral equilibria of noncooperative games. \textit{Theory and Decision}, 78, 171-188.

\medskip

\noindent Nau, R.\ F., McCardle, K.\ F.\ (1990). Coherent behavior in noncooperative games. \textit{Journal of Economic Theory}, 50, 424-444

\medskip

\noindent Nau, R.\ F., McCardle, K.\ F.\ (1991). Arbitrage, rationality, and equilibrium. \textit{Theory and Decision}, 31, 199-240.
\medskip

\noindent Nielsen, M. (2019). On linear aggregation of infinitely many finitely additive probability measures. \textit{Theory and Decision}, 86, 421-436.

\medskip

\noindent Nielsen, M. (2021). The strength of de Finetti’s coherence theorem. \textit{Synthese}, 198, 11713-11724.

\medskip

\noindent Phelps, R.\ R.\ (2001). \emph{Lectures on Choquet’s Theorem}, 2nd ed., New York: Springer.
\medskip

\noindent Radner, R. (1980). Collusive behavior in noncooperative epsilon-equilibria of oligopolists with long but finite lives. \textit{Journal of Economic Theory}, 22, 136-154.

\medskip

\noindent Samet, D. (1998). Common priors and separation of convex sets. \textit{Games and Economic Behavior}, 24, 172-174.
\medskip

\noindent Savage, L.\ J. (1954). \emph{The Foundations of Statistics}, New York: John Wiley \& Sons.

\medskip

\noindent Sion, M. (1958). On general minimax theorems. \textit{Pacific Journal of Mathematics}, 8, 171-176.

\medskip

\noindent Stone, M.\ (1961). The opinion pool. \textit{The Annals of Mathematical Statistics}, 1339-1342.


\end{document}